\documentclass[a4paper,11pt]{article}
\pdfoutput=1
\usepackage{jheppub}
\usepackage[T1]{fontenc}
\usepackage{diagbox}
\usepackage{adjustbox}
\usepackage{multirow}
\usepackage{pifont}
\usepackage{array}
\usepackage{mathtools}
\usepackage{booktabs}
\usepackage[dvipsnames]{xcolor}
\usepackage{tcolorbox}
\usepackage{float}
\usepackage{placeins}
\usepackage{braket}
\usepackage{slashed}
\usepackage{makecell}

\newcommand{\vev}{v_\text{EW}}

\newcommand{\Huayang}[1]{{\color{red} [HS: #1]}}
\newcommand{\Xia}[1]{{\color{Magenta} [Xia: #1]}}

\usepackage[normalem]{ulem}
\newcommand{\stkout}[1]{\ifmmode\text{\sout{\ensuremath{#1}}}\else\sout{#1}\fi}

\newcommand{\changed}[2]{{\protect\color{red}\stkout{#1}}{\protect\color{blue}\uwave{#2}}}

\newcommand{\beq}{\begin{equation}}
\newcommand{\eeq}{\end{equation}}
\newcommand{\bea}{\begin{eqnarray}}
\newcommand{\eea}{\end{eqnarray}}

\newcommand{\la}{\langle}
\newcommand{\ra}{\rangle}
\newcommand{\s}{\sigma }
\newcommand{\cgam}{c_\gamma}
\newcommand{\sgam}{s_\gamma}

\title{Matching the real Higgs triplet extension of Standard Model to HEFT  
}

\author[a]{Huayang Song,}
\author[b]{Xia Wan,}

\affiliation[a]{Particle Theory and Cosmology Group, Center for Theoretical Physics of the Universe, Institute for Basic Science (IBS), Daejeon, 34126, Korea}
\affiliation[b]{School of physics and Information Technology, Shaanxi Normal University,
Xi'an 710119, China}

\emailAdd{huayangs1990@ibs.re.kr}
\emailAdd{wanxia@snnu.edu.cn}

\abstract{
We make a first matching of the real Higgs triplet extension (RHTE) of the standard model to the Higgs effective field theory (HEFT), which is also known as electroweak chiral Lagrangian (EWChL). 
In the RHTE, the ratio $\xi$ of two VEVs 
separately from triplet and doublet is a tiny quantity, due to the constrained custodial symmetry violation. 
Based on the nonlinear representation~\cite{Song:2024kos} of RHTE and functional method, 
we get the HEFT in the power counting of $\xi$.
By comparing its phenomenological results of $hh\to hh$, $W W \to h h$, $Z Z \to hh $ to SMEFT's, we show its convergence property is as good as SMEFT's. Besides we explicitly show that the HEFT is available in the phase transition region where SMEFT fails.}

\begin{document} 
\maketitle
\flushbottom

\section{Introduction\label{section:introduction}}
After the discovery of the Higgs boson, the last piece of the Standard Model (SM), at the Large Hadron Collider (LHC) in 2012~\cite{CMS:2012qbp,ATLAS:2012yve}, the major efforts of the experiments (especially the LHC) has turned to searches for new particles at higher energy scales and precision measurements of Higgs and electroweak (EW) observables. So far no significant deviation from the SM predictions has been observed, which indicates that the potential new physics beyond the SM (BSM), if it exists, must be at a higher energy scale than that probed by the current experiments. In this scenario, effective field theories (EFTs) provide us the general and consistent framework to parameterize the BSM effects and possible deviations of the SM in the future experiments.  

Two types of EFTs are commonly used to model the BSM physics in high-energy physics community - the SM Effective Field Theory (SMEFT)~\cite{Brivio:2017vri} and the Higgs Effective Field Theory (HEFT)~\cite{Appelquist:1980vg, Longhitano:1980iz, Longhitano:1980tm, Feruglio:1992wf, Herrero:1993nc, Herrero:1994iu, Grinstein:2007iv, Buchalla:2012qq, Alonso:2012px, Buchalla:2013rka, Brivio:2013pma, Buchalla:2013eza, Gavela:2014vra, Pich:2015kwa, Alonso:2015fsp, Brivio:2016fzo, Alonso:2016oah, Pich:2016lew, Merlo:2016prs, Pich:2018ltt, Krause:2018cwe, Sun:2022ssa, Sun:2022snw}. Both are constructed by exclusive SM degrees of freedom and invariably under the SM gauge group $SU(3)_c\times SU(2)_L\times U(1)_Y$. However, they treat the Higgs field, $h$ and the electroweak Nambu-Goldstone bosons, $\pi_i,i=1,2,3$ differently. The SMEFT embeds $h$ and $\pi_i$'s in an $SU(2)_L$ scalar doublet while the HEFT separates them into a singlet and a triplet.
Fundamentally the SMEFT describes a Higgs doublet in unbroken phase and its scalar manifold contains an $O(4)$ invariant fixed point~\cite{Alonso:2016oah} while the HEFT generally no longer requires this condition and can be reformulated into a SMEFT if and only there is an $O(4)$ invariant point. Thus HEFT is more general than SMEFT. 

The procedure of connecting UV model to EFT is called matching, which provides a top-down view to study EFT. If there exist non-zero wilson coefficients from experiments, the results of matching would tell which UV model is favored. 
Though the matching between UV models and SMEFT are well considered~\cite{Henning:2014wua, Drozd:2015kva, Chiang:2015ura, Huo:2015exa, Huo:2015nka, Brehmer:2015rna, Crivellin:2016ihg, Belusca-Maito:2016dqe, Dawson:2017vgm, Corbett:2017ieo, deBlas:2017xtg, Han:2017cfr, Jiang:2018pbd, Craig:2019wmo, Haisch:2020ahr, Gherardi:2020det, Dawson:2020oco, Marzocca:2020jze, Corbett:2021eux, Zhang:2021tsq, Zhang:2021jdf, Brivio:2021alv, Dedes:2021abc, Dawson:2021xei, Du:2022vso, Li:2022ipc, Dawson:2022cmu, Zhang:2022osj, Liao:2022cwh, 
Ellis:2023zim, Dawson:2023ebe, Li:2023cwy, Li:2023ohq, DasBakshi:2024krs, Dawson:2024ozw}~\footnote{Due to the rapid development of this area, we apologize for those work that we forget to cite here. Nowadays there are several tools developed for matching UV models to SMEFTs automatically: \textbf{MatchingTools}~\cite{Criado:2017khh} is a Python library for addressing EFT matching at tree level using functional
approach, \textbf{CoDEx}~\cite{DasBakshi:2018vni} and \textbf{Matchete}~\cite{Fuentes-Martin:2022jrf} are  Mathematica packages performing EFT matching up to 1-loop level utilizing functional methods, and \textbf{Matchmakereft}~\cite{Carmona:2021xtq} is a tool that automates the tree-level and one-loop matching via the plain Feynman diagram approach}, the matching to HEFT is less performed and only more commonly considered in composite Higgs mdoels (CHMs)~\cite{Grojean:2013qca, Alonso:2014wta, Hierro:2015nna, Gavela:2016vte, Qi:2019ocx, Lindner:2022kxm}. 
A systematic study firstly appears in 
Ref.~\cite{Buchalla:2016bse}, which matches the SM model with an extra singlet to linear and nonlinear EFTs and clarify their relations. A recent study of matching the SM with real singlet extension, complex singlet extension and 2 Higgs doublet model~(2HDM) to HEFT in  Ref.~\cite{Dawson:2023oce}, shows that different power countings (PCs) in HEFT can yield different results. Ref.~\cite{Banta:2023prj} proposed to use ``straight-line'' basis of the 2HDM instead of \textit{Higgs basis} as in Refs.~\cite{Dawson:2023ebe} and obtained a SMEFT-like EFT with better convergence property. Refs.~\cite{Arco:2023sac, Buchalla:2023hqk} match 2HDM to HEFT via diagrammatic method and functional method respectively and focus on the non-decoupling effects from heavy states. 
In this work, we make the first matching of the SM with one real Higgs triplet extension (RHTE) to HEFT through functional method~\footnote{Matching RHTE to EFT in broken phase is actually performed before in \cite{Chivukula:2007koj, Cohen:2020xca}, however, their EFTs are not HEFT as in \cite{Brivio:2016fzo, Sun:2022ssa}. Especially its Goldstone modes are not in $U(\pi)$ matrix.}. 


Unlike the previous studied models, the extra scalar state in the RHTE can develop a non-vanishing vacuum expectation value (VEV) which also breaks the electroweak symmetry and contributes to the value of the electroweak VEV $v_\text{EW}$~\footnote{Generally both scalar doublets in the 2HDM can develop non-vanishing VEVs, however one can perform a rotation into \textit{Higgs basis} and only one doublet contributes to the EW VEV.}. In such a case, one can no longer embed the Goldstones and part of the physical Higgs particle into a doublet, which superficially indicates that the HEFT is a more suitable EFT to describe such kind of model like the RHTE. Due to the low energy measurements, the value of the extra VEV, $v'$, is constrained to be small. In the matching process, one usually assume that the masses of the BSM states are larger enough to perform a series expansion. In a model like RHTE, the large values of masses typically indicate the smallness of the ratio $v'/v$. However, one should note that this is not a ``necessary and sufficient'' condition. Therefore compared to a large mass assumption, the small ratio $v'/v$ provides us a more natural expansion parameter and we would expect that it could delineate a larger region of parameter space.

The paper is organized as follows: we start by recapping the HEFT in section~\ref{sec:HEFT}, then we introduce a nonlinear representation of RHTE in section~\ref{sec:RHTE}.
We next perform a matching of the RHTE to the HEFT through functional method in section~\ref{sec:mat2HEFT}, at tree level and partly at one-loop level.  Section~\ref{sec:mat2SMEFT} is devoted to present the results from both the HEFT and the SMEFT matchings, and discuss their performance.
Finally in section~\ref{sec:concl} we present our conclusions.


\section{HEFT~\label{sec:HEFT}}
As discussed in the introduction, the HEFT treats the Higgs field and Golstone fields differently. Following the Callan-Coleman-Wess-Zumino (CCWZ) construction~\cite{Coleman:1969sm}, the EW would-be Goldstones $\pi_i$ are embeded into a dynamical unitray matrix $U\equiv\exp\left(\frac{i \pi_i \sigma_i}{v}\right)$, where $\sigma_i$ are the three Pauli matrices. Inspired by the similarity between the EW Goldstone bosons and pions, the HEFT operators can be constructed order by order in the number of the covariant derivatives (therefore the HEFT is also known as the electroweak chiral Lagrangian with a light Higgs boson). At the lowest order, the HEFT Lagrangian is given as,
\begin{align}
\mathcal{L}_{\text{HEFT}}^{\text{LO}}\supset&\frac{1}{2}\mathcal{K}(h)D_\mu h D^\mu h-\mathcal{V}(h)-\frac{\vev^2}{4} \mathcal{F}(h)\la V_\mu V^\mu \ra 
+\frac{\vev^2}{4}\mathcal{G}(h)\la V_\mu\sigma_3 \ra \la V^\mu\sigma_3 \ra \nonumber \\
&-\frac{\vev}{\sqrt{2}}(\bar{Q}_L U\mathcal{Y}_Q(h) Q_R+\bar{L}_L U\mathcal{Y}_L(h)L_R + h.c.)~,
\label{eq:HEFT}
\end{align}
where $D_\mu$ is the covariant derivative, $\vev=246$~GeV represents the vacuum expectation value (VEV) of the Higgs field, $\la ... \ra$ denotes the trace, $V_\mu\equiv U^\dagger D_\mu U$ is a shorter notation, and $\mathcal{K}(h)$, $\mathcal{V}(h)$, $\mathcal{F}(h)$, $\mathcal{G}(h)$, and $\mathcal{Y}_{Q/L}(h)$ are polynomial functions of $h$ with the following forms,
\bea
    \mathcal{K}(h)&=&1+c_1^k\frac{h}{\vev}+c_2^k\frac{h^2}{\vev^2}+\cdots, \\
    \mathcal{V}(h)&=&\frac{1}{2}m_{h}^2 h^2\left[1+(1+\Delta\kappa_3)\frac{h}{\vev}+\frac{1}{4}(1+\Delta\kappa_4)\frac{h^2}{\vev^2}+\cdots\right],  \\
    \mathcal{F}(h)&=&1+2(1+\Delta a)\frac{h}{\vev}+(1+\Delta b)\frac{h^2}{\vev^2}+\cdots, \\
    \mathcal{G}(h)&=&\Delta\alpha+\Delta a^\slashed{C}\frac{h}{\vev}+\Delta b^\slashed{C}\frac{h^2}{\vev^2}+\cdots, \\
    \mathcal{Y}_Q(h)&=&\text{diag}\left(\sum_n Y_U^{(n)}\frac{h}{\vev^n},\,\sum_n Y_D^{(n)}\frac{h^n}{\vev^n}\right), \nonumber\\ \mathcal{Y}_L(h)&=&\text{diag}\left(0,\,\sum_n Y_\ell^{(n)}\frac{h^n}{\vev^n}\right).
\eea
The normalization function $\mathcal{K}(h)$ of the Higgs kinetic term generally is redundant. One can perform a field redefinition $h'\rightarrow\int_0^{h}\sqrt{\mathcal{K}(s)}ds$, and reabsorb its entire effects into redefinitions of the other coefficients. But we observe that such a form of the Higgs kinetic normalization function is commonly generated in the middle step of EFT matching. The matching results can be expressed in a more compact way with it. Therefore in the following we use such an over-complete basis without performing the redefinition but only present the Wilson coefficients in the non-redundant basis in the appendix. The second term contains the Higgs potential; the third term describes the kinetic terms of the gauge bosons; the fifth term accounts for the Yukawa interactions in which the $n=0$ terms yield fermion masses. In this work we will focus on the bosonic sector, therefore we will not consider the $\mathcal{Y}_{Q, L}$ terms anymore, which does not impact our results.

Now, let us comment on the fourth term, which violates the custodial symmetry. It is usually considered as a next-to-leading order (NLO) term (a chiral dimension 4 term). This is inspired by observing that the custodial violation effect is at least generated at the one-loop level (dominantly by the top quark) in the SM. With only the common chiral power-counting scheme, this treatment is based on an assumption of the UV theory that there exists an approximate remnant symmetry (custodial symmetry) after the electroweak symmetry breaking in the UV model. Although from the experimental measurements we know that the custodial symmetry violation is rather small, its origin is still unclear. It is definitely possible that the custodial violation is generated by some loop effects like in the SM, but it can also be triggered by some tree-level new physics, especially at the electroweak scale. This commonly happens in the BSM models with an extended scalar sector. Just like in the RHTE that will be discussed, the custodial symmetry breaking effects are generated at the same time when the electroweak symmetry is broken. Even though the sizes of custodial symmetry-preserving and -violating effects have significant differences, from a low-energy point of view it is no reason to assume that the $\braket{V_\mu\sigma_3}\braket{V^\mu\sigma}$ operator is a chiral dimension-four operator, which is considered loop-generated. It is also unnatural to relate it to some scales considerably higher than the electroweak scale since it turns out at the same time with other custodial symmetry-preserving terms during spontaneous symmetry breaking. This inspires that Ref.~\cite{Sun:2022ssa} put this term at the leading order (LO) when they only adapt the standard chiral dimension counting. Here we also use this convention.

From the above discussion, one can notice that, given the observed smallness of the custodial symmetry violation, a power counting rule with two expansion parameters is more systematic. In addition to the standard chiral dimension (roughly speaking, the external momentum $p$), a parameter $\xi$, which we use the ratio between the VEVs of two scalars in later discussion, characterizes the size of the custodial symmetry-breaking effect. This is not something new. In composite Higgs models (CHMs), there exist two scales beyond the electroweak scale, a global symmetry breaking scale $f$ and a cutoff scale $\Lambda_s=g_c f$, where the low-energy description of is cut off and $g_c$ denotes the strong coupling constant~\cite{Buchalla:2014eca}. With naive dimensional analysis (NDA)~\cite{Manohar:1983md}, one can relate the chiral dimensions to the number of loops by choosing $g_c\simeq 4\pi$ and the resulting EFT operators can be conveniently organized in the standard HEFT framework. Or in the limit $\vev/f\rightarrow 0$, the new physics completely decouples and the SM is recovered, with a SMEFT description. But residue effects of finite $\vev/f$ are unavoidable in CHMs. Generally without any assumptions or knowledge of the values of $g_c$, the double expansion in powers of $\vev/f$ (known as the vacuum misalignment angle) and $f/\Lambda_s=1/g_c$ should be considered at the same time~\footnote{One should notice that even one takes $f^2/\Lambda^2=1/(16\pi^2)$, the double expansion in the number of loops and in powers of $\vev/f$ should also be performed in the EFT~\cite{Buchalla:2014eca}.}. This is the key feature when matching a UV theory with multiple scales to an EFT, especially the physics between the two UV scales is hard to describe separately. In the following, we match the RHTE to the HEFT in powers of $\xi$ by integrating out the BSM states at the tree level~\footnote{Note that the loop expansion here is not the number of loops in the EFT, which is related to the chiral dimension, but the number of loops in the UV calculation.}. One should note that the chiral dimension provides a systematic way to organize the HEFT operators but it is generally impossible to perform a matching of the HEFT to a UV complete model without introducing an extra expansion parameter. 

\section{The SM with one real Higgs triplet extension (RHTE)~\label{sec:RHTE}}
\subsection{The triplet in nonlinear form}
In the RHTE besides the $SU(2)_L$ doublet there exists a triplet with hyper-charge $Y=0$. 
The Lagrangian of RHTE in scalar sector can be expressed as~\cite{Corbett:2021eux,Ellis:2023zim}
\bea
{\cal L}&=& \left(D_{\mu} H  \right)^{\dagger}\left(D^\mu  H \right)+\langle D_{\mu}\Sigma^{\dagger} D^{\mu}\Sigma\rangle -V,
\label{LHSigma}\\
V&=&
Y_1^2   H^\dagger  H +Z_1  ( H^\dagger  H)^2
+ Y_2^2 \langle \Sigma^\dagger \Sigma \rangle
+  Z_2 \langle \Sigma^\dagger \Sigma \rangle^2 
+ Z_3  H^\dagger  H \langle \Sigma^\dagger\Sigma \rangle
+2 Y_3   H^\dagger \Sigma H,
\label{Hpotential}
\eea
where $\langle ... \rangle$ denotes the trace, 
$Y_is,i=1,2,3$ are dimensional parameters while $Z_is,i=1,2,3$ are dimensionless. $ H $ is  
the SM Higgs doublet and $\Sigma$ is the real triplet. After spontaneously symmetry breaking (SSB)
the doublet is 
\begin{equation}
H = U \frac{1}{\sqrt{2}}
\left(\begin{array}{c}
 0 \\
 v+h^0\\
\end{array} \right)~,\quad U\equiv \exp\left(\frac{i \pi_i \sigma_i}{v}\right)
\label{eq:H}
\end{equation}
where $\sigma_i,i= 1,2,3$ are Pauli matrices, $v$ is vacuum expectation value (VEV) of the doublet, $h^0$ is the radial mode associated to the VEV direction, and the unitary matrix $U$ encapsulate other three degrees of freedom $\pi_i$. In SM $\pi_i$s are Goldstones while in RHTE they are generally not. 
We write $H$ in ``polar coordinates'' instead of ``Cartesian coordinate'' so that it is near the nonlinear form in HEFT.   


Similarly 
we define a nonlinear ``rotated'' 
triplet, 
which is    
\begin{equation}
\Sigma =
U \Phi
U^{\dagger}~, \quad \Phi\equiv \phi_i \sigma_i/2=\frac{1}{2}\left(\begin{array}{cc}
 v^\prime +\phi^0 & \phi_1 -i \phi_2  \\
 \phi_1 + i \phi_2 &   -v^\prime -\phi^0
\end{array} \right),
\label{eq:SigmaPhi}
\end{equation}
where $\Sigma$ is related to $\Phi$ through the $U$ matrix,
$v^\prime$ is the VEV of the triplet field, and $\phi^0=-v^\prime+\phi_3$ is the corresponding radial mode. As the $U$ matrix could be considered as an $SU(2)_L$ rotation with ``angles'' of $\pi_i/(2v)$, in Eq.~\eqref{eq:SigmaPhi} we rotate $\Phi$ to $\Sigma$ by ``angles'' of $\pi_i/(2v)$.
This rotation does not change physical features such as electric charge and VEV, i.e. $\la \Sigma \ra = \la \Phi\ra$. 
Inserting Eq.s~\eqref{eq:H} and~\eqref{eq:SigmaPhi} back into the potential in Eq.~\eqref{Hpotential}, it is noticeable that the $U$ matrix disappears completely as $U$s cancel out in each term of $V$. This means there is no mass mixing between $\pi_i$s and $\phi_i$s. However, next we will see the mixing between $\pi_i$s and $\phi_i$s appear in kinetic terms. 

From definition the real triplet $\Sigma$ transforms under $SU(2)_L\times U(1)_Y$ as 
\begin{align}
    \Sigma\to \mathfrak{g}_L(\boldsymbol{\theta}) 
    ~\Sigma ~ \mathfrak{g}_L(\boldsymbol{\theta})^\dagger~,  \quad \mathfrak{g}_L(\boldsymbol{\theta})= { \rm exp}(\frac{i}{2} \theta_i \sigma_i).
\label{gL}
\end{align}
Since the $U$ matrix transforms as~\cite{Sun:2022ssa} 
\bea
   U\to \mathfrak{g}_L(\boldsymbol{\theta}) 
   ~U~ \mathfrak{g}_Y(\theta_Y)^\dagger, \quad
   \mathfrak{g}_Y(\theta_Y)=
   \rm{exp}(\frac{i}{2}
   \theta_Y \sigma_3), \label{Utransformation}
\label{gY}
\eea
$\Phi$ should transform as 
\bea
   \Phi\to \mathfrak{g}_Y(\theta_Y) 
    ~\Phi ~ \mathfrak{g}_Y(\theta_Y)^\dagger.
\label{PhiTransformation}
\eea
 Obviously $\Phi$ is not a $SU(2)_L$ triplet. Correspondingly, 
 the covariant derivatives are
 \bea
    D_\mu U &=& \partial_\mu U
    + i  \hat{W}_\mu U -i  U \hat{B}_\mu, \quad  \hat{W}_\mu= g W_{i \mu } \sigma_i/2, \hat{B}_\mu= g^\prime B_\mu \sigma_3/2 \label{eq:WmuBmu}\\
    D_\mu \Phi &=& \partial_\mu \Phi+
    i g^\prime [\hat{B}_\mu, \Phi].
\eea


\subsection{Kinetic mixing}

The Lagrangian of kinetic terms is
 \bea
 \mathcal{L}_{\rm kin}& = &
 D_\mu H^\dagger D^\mu H + \la D_\mu \Sigma^\dagger D^\mu \Sigma \ra~,
 \eea
Inserting Eqs.~\eqref{eq:H} and \eqref{eq:SigmaPhi} there appears a kinetic mixing term from $\la D_\mu \Sigma^\dagger D^\mu \Sigma \ra$, 

\bea
\mathcal{L}_{\rm mix}= -v^\prime \epsilon_{3jk}D_\mu \phi_j 
 D^\mu \pi_k/v,
 \eea 
where $j/k=1,2,3$, $D_\mu \phi_j =1/2 \la D_\mu \Phi \sigma_j \ra$, $D_\mu \pi_k$ comes from  $U^\dagger D_\mu U$, i.e. $D_\mu \pi_k \subset 1/2\la U^\dagger D_\mu U \sigma_k \ra$. $j=3$ or $k=3$ would make $\mathcal{L}_{\rm mix}$ absent, which indicates the neutral part of $\Phi$, $\phi_3$ (or $\phi^0$) does not involve in the mixing, only the charged part $\phi_1$/$\phi_2$ contributes to Goldstones. Meanwhile, 
$\mathcal{L}_{\rm mix}$ is proportional to $v^\prime$, which means if the triplet VEV becomes zero ($v^\prime \to 0$), no mixing appears and $\pi_i$s become Goldtones as in SM. Anyhow, let us diagonalize this mixing term and see what happens.  

Using $\phi^{\pm}\equiv (\phi_1 \mp i \phi_2)/\sqrt{2}$ and $\pi^{\pm}\equiv (\pi_1 \mp i \pi_2)/\sqrt{2}$, we rewrite the mixing term as 
\bea
\mathcal{L}_{\rm mix}&=&
\left(\begin{array}{c c}
D_\mu \phi^+  &  -i \sqrt{1+ 4 \xi^2} v D_\mu \pi^+
\end{array}
\right)
\left(\begin{array}{c c }
  1  & \frac{2\xi}{\sqrt{1+4 \xi^2}}  \\
   \frac{2\xi}{\sqrt{1+4 \xi^2}} &  1 \\
\end{array} \right)
 \left(\begin{array}{c }
   D_\mu \phi^- \\
    i \sqrt{1+ 4 \xi^2} v D_\mu \pi^-\\
\end{array} \right)~,
\eea
where we define $\xi\equiv v^\prime/v$. This kinetic matrix can be diagonalized by rotating $\phi^{ \pm}$ and 
$\pi^{ \pm}$ fields through
\bea
\left(\begin{array}{c }
    \phi^- \\
    i  \sqrt{1+ 4 \xi^2} v \pi^-\\
\end{array} \right)
=
\left(\begin{array}{c c }
   \sqrt{1+4\xi^2} & \quad 0  \\
   -2\xi & \quad  1 \\
\end{array} \right)
\left(\begin{array}{c }
   \phi^{\prime -} \\
    i  \sqrt{1+ 4 \xi^2}  v \pi^{\prime -}\\
\end{array} \right)~,
\label{Hmix}
\eea
where $\phi^{\prime\pm}$ represent heavy charged Higgs  and $\pi^{\prime \pm}$ are massless Goldstones, $\xi$ is a small quantity of $\mathcal{O}(10^{-2})$ or less~\cite{Cheng:2022hbo}.
The difference of $\pi^{\pm}$ and 
$\pi^{\prime \pm}$ could cause a difference of order $\mathcal{O}(\xi)$ between $U=
\exp{\frac{i \pi_i \sigma_i}{v}}$ and $U^\prime=
\exp{\frac{i \pi^\prime_i \sigma_i}{v}}$, which is

\bea
U &=&  \exp \left (
\frac{i\sigma_i \pi^\prime_i}{v}+
\frac{2\xi}{\sqrt{1+4\xi^2}} i (-\sigma_1 \phi^\prime_2+\sigma_2 \phi^\prime_1)\right)\nonumber\\
&=& U^\prime+ \xi M_1(\pi^\prime,\phi^{\prime\pm})+ {\xi^2}M_2(\pi^\prime,\phi^{\prime\pm})+...
\label{eq:Uprime}
\eea
where $\phi^{\prime \pm}=(\phi^\prime_1\mp i \phi^\prime_2)/\sqrt{2} $,
the matrices $M_1(\pi^\prime,\phi^{\prime\pm})$ and $M_2(\pi^\prime,\phi^{\prime\pm})$ are proportional to $\phi^{\prime\pm}$ fields, 
the ellipsis represents higher orders.

From Eq.~\eqref{eq:Uprime} we see $U$ depends not only on $U^\prime$ but also on $\pi^\prime_i$s,  
so $U^\prime$ and $\pi^\prime$ would coexist in the Lagrangian of UV model. Encapsulating Goldstones in a unitary matrix become complicated. All in all, we make a dilemma while we meanwhile write the Higgs doublet as $U$ multiplying $(0~~v+h^0)^T$  and let triplet have a non-zero VEV $v^\prime$. The $U$ matrix associated with the doublet VEV $v$ must not be Goldstones while there appear new VEVs.

If we do not use a ``ratoted'' triplet as in Eq.~\eqref{eq:SigmaPhi} but write triplet in linear form as
\begin{align}
\Sigma=\frac{1}{2} \Sigma_i \sigma_i
=
\frac{1}{2} 
\begin{pmatrix}
v_\Sigma+\Sigma^0  & \sqrt{2}\Sigma^+\\
\sqrt{2}\Sigma^- & - v_\Sigma-\Sigma^0 
\end{pmatrix}, 
\label{Sigmamatrix}
\end{align}
there exists no kinetic mixing between $\pi_i$ and $\Sigma_i$. Nevertheless, 
there appears mass mixing between them (see Eq.~\eqref{Hpotential}), which also means $\pi_i$ in $U$ are not Goldstones if the triplet VEV is not zero. This linear form is also not a good choice for HEFT matching. 


\subsection{The nonlinear representation }


A feasible solution is to reform the doublet. Just imagine that in Eq.~\eqref{eq:H}
$U$ is ``perpendicular''  to the radial mode $v+ h^0$, which also says the Goldstones $U^\prime$ are not
``perpendicular'' to the ``direction'' of $v+ h^0$, but some new ``direction''. Thus we reform the doublet as 

\begin{equation}
H =  U^\prime 
\left(\begin{array}{c}
 \chi^+ \\
 \frac{1}{\sqrt{2}}(v+h^0+i \chi^0)\\
\end{array} \right)~,
\label{eq:H_Uprime}
\end{equation}
where $(\chi^+ \quad \frac{1}{\sqrt{2}}(v+h^0+i\chi^0))^T$ represents a new ``direction'', 
$\chi^\pm$/$\chi^0$ are some physical states. Correspondingly, the triplet are rotated by $U^\prime$ instead of $U$, which is 
\begin{equation}
    \Sigma =
U^\prime \Phi^\prime
U^{\prime \dagger}~, \quad \Phi^\prime\equiv \phi^\prime_i \sigma_i/2, \quad \phi^{\prime\pm} = \frac{1}{\sqrt{2}} \phi^\prime_1\mp i \phi^\prime_2~.
\label{eq:Sigma_Uprime}
\end{equation}
The left problem is what are $\chi^\pm$/$\chi^0$? 
First they have to make kinetic mixing or mass mixing disappear.  
Second they must be some functions of $\phi_i$s since they are ``perpendicular'' to both the Goldstones and $v+h^0$ ``direction''s. While expanding $D_\mu H^\dagger D^\mu H $  with Eq.\eqref{eq:H_Uprime}, we find a new kinetic mixing term 
\bea
\mathcal{L}^H_{\rm mix}= \epsilon_{3jk}D_\mu \chi_j 
 D^\mu \pi^\prime_k/2
 \eea 
appear, compare it with the kinetic mixing term from $\la D_\mu \Sigma^\dagger D^\mu \Sigma \ra $,  
\bea
\mathcal{L}^\Sigma_{\rm mix}= -v^\prime \epsilon_{3jk}D_\mu \phi^\prime_j 
 D^\mu \pi^\prime_k/v,
 \eea 
we get 
\beq
\chi^\pm = 2 \frac{v^\prime}{v} \phi^{\prime\pm}, \quad \chi^0 = 0,  
\eeq
 which is the right ``direction''. Since $\chi^\pm$ is proportional to $v^\prime/v$, when $v^\prime$ become zero, the doublet recover to its SM form as in Eq.~\eqref{eq:H}.

 Under this representation, we only have $U^\prime$, $h^0$ and $\phi^\prime_i$s in the RHTE model, integrating heavy particles will not destroy $U^\prime$ matrix. Thus, matching it to HEFT through functional method become straightforward and feasible. In the following we will use $U$ and $\phi_i$s back to denote Goldstones and physical states since we do not need to redefine fields any more. This nonlinear representation could be easily extended to new physics model with general scalar extensions, e.g., complex triplet, quadruplet, and septet~\cite{Song:2024kos}.

\subsection{Masses of heavy scalars}

After minimizing the potential in 
Eq.~\eqref{Hpotential}, 
we have the following relations
\beq
Y_1^2= - Z_1 v^2 - Z_3 v^{\prime 2}/2 + Y_3 v^{\prime}, \quad 
Y_2^2 = -Z_3 v^2/2 - Z_2  v^{\prime 2} + \frac{Y_3 v^2}{ 2 v^{\prime}}.
\label{Y1Y2}
\eeq
So instead of using the parameter set of $(Z_1, Z_2, Z_3, Y_3, Y_1, Y_2)$, we have a new parameter set of $(Z_1, Z_2, Z_3, Y_3, v, v^\prime)$. With the non-zero $v$ and $v^\prime$, after symmetry breaking, mass terms are
\begin{align}
\mathcal{L}_{\rm mass}=\frac{1}{2}\begin{pmatrix}
    h^0 & \phi^0
\end{pmatrix}\begin{pmatrix}
    2Z_1 v^2 & v\left(Z_3 v'-Y_3\right) \\
    v\left(Z_3 v'-Y_3\right) & 2Z_2 v'^2+\frac{Y_3 v^2}{2v'}
\end{pmatrix}\begin{pmatrix}
    h^0 \\ \phi^0
\end{pmatrix}+\left(v^2+4v'^2\right)^2\frac{Y_3}{2v' v^2}\phi^+\phi^- \label{eq:massmatrix}
\end{align}
where $h^0$ and $\phi^0$ can be rotated into the mass eigenstates $h$ and $K$ via an angle $\gamma$
\beq
\left(\begin{array}{c}
 h \\
 K \\
\end{array} \right)
= 
\left(\begin{array}{cc}
 \cos \gamma & -\sin \gamma \\
 \sin \gamma & \cos \gamma 
\end{array} \right)
\left(\begin{array}{c }
 h^0 \\
 \phi^0\\
\end{array} \right).
\label{eq:NSDiagMatrix}
\eeq
The masses of $h$ and $K $ are 
\begin{align}
m_{h, K}^2=Z_1 v^2+Z_2 v^{\prime 2}+\frac{Y_3 v^2}{4 v^{\prime}}\mp\sqrt{\left(Z_1 v^2-Z_2 v^{\prime 2}-\frac{Y_3 v^2}{4 v^{\prime}}\right)^2+v^2 (Z_3 v^{\prime}-Y_3)^2}. 
\label{eq:NSmass2}
\end{align}
The lighter one $h$ is the discovered 125 GeV Higgs, and $K$ is a heavy neutral scalar.
$\phi^\pm$ are charged Higgs, 
after normalizing their kinetic term, $(1+ 4(v^\prime/v)^2 ) D_\mu \phi^+ D^\mu \phi^-$,
from the second term in Eq.\eqref{eq:massmatrix} we get their masses as 
\beq
m_{\phi^\pm}^2=(v^2+4 v^{\prime 2})\frac{Y_3 }{2 v^\prime}.
\label{eq:mHpm}
\eeq

\subsection{Theoretical constraints on the parameters}
From Eq.~\eqref{eq:mHpm}  we learn that to make the charged Higgs mass squared positive implies $ Y_3/v' >0$. With our phase convention $v'>0$, we choose $Y_3>0$ without loss of generality. In principle, one can express the Lagrangian parameters $Y_3$ and the $Z$'s in terms of the physical Higgs masses and the mixing angle $\gamma$ as well as the VEV's $v$ and $v'$.
\begin{align}
    Y_3&=\frac{2v'}{v^2+4v'^2}m_{\phi^\pm}^2 \\
    Z_1&=\frac{1}{2v^2}\left(c_\gamma^2 m_{h}^2+s_\gamma^2 m_{K}^2\right) \\
    Z_2&=\frac{1}{2v'^2}\left(s_\gamma^2 m_{h}^2+c_\gamma^2 m_{K}^2-\frac{v^2}{v^2+4v'^2}m_{\phi^\pm}^2\right) \\
    Z_3&=\frac{1}{v v'}\left[s_\gamma c_\gamma\left(m_{h}^2-m_{K}^2\right)+\frac{2v v'}{v^2+4v'^2}m_{\phi^\pm}^2\right] 
\end{align}
The tachyonless condition in the CP even scalar sector reads (mass matrix is positive-definite)
\begin{align}
    -Y_3^2 v'+Y_3(Z_1 v^2+2Z_3 v'^2)+(4Z_1 Z_2-Z_3^2)v'^3>0&\qquad(\text{determinant is positive}) \\
    Y_3 v^2+4Z_1 v^2 v'+4Z_2 v'^3>0&\qquad(\text{trace is positive})
\end{align}
Given the boundness from below conditions for the potential
\begin{align}
    Z_1,\, Z_2\geq 0,\quad |Z_3|\geq -2\sqrt{Z_1 Z_2}
\end{align}
the second equation of the tachyonless condition is automatically satisfied. The first equation leads to constraints on $Y_3$
\begin{align}
    \max(0, Y_3^-)<Y_3<Y_3^+
\end{align}
with the expressions of $Y_3^\pm$ being given by
\begin{align}
    Y_3^\pm=\frac{1}{2v'}\left(Z_1 v^2+2Z_3 v'^2\pm\sqrt{Z_1^2 v^4+4 Z_1 Z_3 v^2 v'^2+16Z_1 Z_2 v'^4}
    \right)
\end{align}
from which we can see that $Y_3$ can not be arbitrarily large as long as we require that the quartic couplings $Z$'s are perturbative.

The model parameter space can also be constrained by requiring perturbative unitarity in scattering processes, which is discussed in Ref.~\cite{Khan:2016sxm, Krauss:2018orw, Ashanujjaman:2024lnr}. Roughly speaking, this requirement imposes the following conditions $Z_{1, 2, 3}\lesssim 4\pi$. However, one can generally consider the RHTE as an effective model describing the intermediate physics between some more fundamental UV physics and the SM, so they can be violated in some level, and we discuss and impose them in this work.

\section{Matching RHTE to the HEFT~\label{sec:mat2HEFT}}

\subsection{Tree level}

The matching procedure at tree level by functional method is, 
first choosing a set of independent parameters, then getting the equation of motions (EoMs) of heavy states, at last solving them in expansion of one parameter or more parameters.
In this work we use the ratio between two VEVs $\xi\equiv v'/v$ as the expansion parameter, due to 
it is a small quantity by experimental constraints of the $\rho$ parameter~\cite{ParticleDataGroup:2024cfk}.

One natural choice of the parameter set is the physical Higgs masses and the mixing angle $\gamma$ as well as the VEV's $v$ and $v'$. However, this cannot show the perturbativity of the quartic couplings explicitly. We choose a parameter set 
\beq
(Z_1,Z_2,Z_3,Y_3,v,\xi)
\eeq
to start the matching. Among them except for $\xi$, others are all at $\mathcal{O}(\xi^0)$ order. This actually make $Y^2_2$, $m^2_{\phi^\pm}$ and $m^2_{K}$ scale as 
\begin{equation}
    Y_2^2 \propto  \mathcal{O}(\xi^{-1}),  \quad m^2_{\phi^\pm}\sim m^2_{K} \propto  \mathcal{O}(\xi^{-1})
    \label{Y22m2}
\end{equation} according to Eqs.~\eqref{Y1Y2}, ~\eqref{eq:NSmass2}
and ~\eqref{eq:mHpm}.
This kind of scaling corresponds to a decoupling limit. That is, if $m_{\phi^\pm}$ and $ m_{K}$ approach infinity (equivalently $\xi$ goes to zero), RHTE degenerates to SM.

This parameter set is good for calculation since they are almost the original parameters in potential (see Eq.~\eqref{Hpotential}). Meanwhile, experimental constraints such as the SM-like Higgs mass $m_h=125$GeV and the electroweak VEV $v_{\rm EW}=v \sqrt{1+4 \xi^2}=246$GeV  are also easy to be imposed. That is, if needed we could covert $(Z_1,Z_2,Z_3,Y_3,v,\xi)$ to $(m_h,Z_2,Z_3,Y_3,v_{\rm EW},\xi)$
by 
\bea
  Z_1 &\to& \frac{m^2_h}{2 v^2}+ \frac{\xi Y_3}{v}+\frac{2 \xi^2(m^2_h-Z_3 v^2)}{v^2} +\frac{\xi^3 (2 m_h^2-Z_3 v^2)^2}{Y_3 v^3}+...,\label{Z1p}\\
  v&\to& v_{\rm EW} /\sqrt{1+4 \xi^2}
\label{vEW}
\eea
where the ellipsis represents higher orders. This replacement does not break the $\xi$ scaling as in Eq.~\eqref{Y22m2}, thus the conversion is trivial.

The Lagrangian of RHTE is written as 
\begin{equation}
\mathcal{L}= \mathcal{L}_{\rm kin }(\phi_1,\phi_2,K,h,U) - V(Z_1,Z_2,Z_3,Y_3,v,\xi; \phi_1,\phi_2,K ,h,U),   
\end{equation}
where $\phi_1,\phi_2,K$ represent heavy Higgses that will be integrated out. We get their EoMs, 
\beq
\partial_\mu\left[ \frac{\partial \mathcal{L}}{\partial (\partial_\mu H^a)} \right] -\frac{\partial \mathcal{L}}{\partial H^a}=0~, \quad H^a=(K,\phi_1,\phi_2).
\label{eq:EoMs}
\eeq
Then we expand the heavy scalars according to $\xi$~\footnote{A general solution should be a Laurent series that include negative exponents. But as a quantum field we do not want a divergent field configuration around $\xi=0$.} as
\bea
K  &= & K_{0} + \xi K_{1} + \xi^2 K_{2}+... \nonumber \\
\phi_1 &=& \phi_{10} + \xi \phi_{11} + \xi^2 \phi_{12}+... \nonumber \\
\phi_2 &=& \phi_{20} + \xi \phi_{21} + \xi^2 \phi_{22}+...,
\label{expandxi}
\eea
and solve Eqs.~\eqref{eq:EoMs} order by order. 

The solutions are
\bea
K_{0}&=&0, \quad K_{1}=\frac{{h}^2}{v}, \nonumber \\
K_{2} &=& \frac{1}{Y_3 v^2} \left[ -2 (h+ v)^2  \la V_\mu  V^\mu \ra  +2 (h+v)^2 \la V_\mu  \sigma_3 \ra \la V^\mu  \sigma_3 \ra -4 D_\mu h  D^\mu h\right. \nonumber \\
&&\left. + (4 Z_1- Z_3)h^2  (h^2+4 h v +5 v^2)  \right], \\
\phi_{10}&=&\phi_{20}=\phi_{11}=\phi_{21}=0, \nonumber\\
\phi_{12}&=&\frac{1}{Y_3 v^2} \left[ 2 (h+v)^2 \la V_\mu \sigma_1 \ra \la V^\mu \sigma_3 \ra - 4 i (2 h-7 v)D_\mu h \la V^\mu \sigma_2 \ra \right. \nonumber \\ 
&&\left. -2 i (h^2-7 h v-8 v^2) \la D_\mu V^\mu \sigma_2 \ra \right], \nonumber \\
\phi_{22}&=&\frac{1}{Y_3 v^2} \left[ 2 (h+v)^2 \la V_\mu  \sigma_2 \ra \la V^\mu  \sigma_3 \ra + 4 i (2 h-7 v) D_\mu h \la V^\mu \sigma_1 \ra \right. \nonumber \\ 
&&\left. +2 i (h^2-7 h v-8 v^2) \la D_\mu V^\mu  \sigma_1 \ra \right],
\label{H1H2}
\eea
where $K$ starts from $\mathcal{O}(\xi)$ while $\phi_{1,2}$ start from  $\mathcal{O}(\xi^2)$, $K$ is gauge invariant, $\phi^\pm= (\phi_1\mp i \phi_2)/\sqrt{2} $ would get a phase factor of $e^{\pm i\theta_Y}$ under gauge transformation of $\mathfrak{g}_L(\boldsymbol{\theta})$ and $\mathfrak{g}_Y(\theta_Y)$ (see Eqs.~\eqref{gL}, \eqref{gY} for the definitions of $\mathfrak{g}_L(\boldsymbol{\theta})$ and $\mathfrak{g}_Y(\theta_Y)$ ) such that the Lagrangian is still gauge invariant. 

With these solutions, we obtain the effective Lagrangian, which reads in ascending powers of $\xi$ up to the terms in $\xi^3$
\bea
\mathcal{L}(\xi^0)&=&\frac{1}{2}D_\mu h D^\mu h + \frac{1}{4}v^4 Z_1 - h^2 v^2 Z_1 - h^3 v Z_1 - \frac{1}{4}h^4 Z_1 - \frac{1}{4}\left(v^2 + 2h v + h^2\right)\la V_\mu V^\mu \ra \nonumber \\
\mathcal{L}(\xi^1)&=&\frac{\xi Y_3}{4v}\left(-v^4 + 4h^2 v^2 + 4h^3 v + h^4\right)\\
\mathcal{L}(\xi^2)&=&\frac{\xi^2}{4 v^2} \left\{8 h^2 D_\mu h D^\mu h + v^6 Z_3 + 8h^2 v^4(2Z_1-Z_3) + 8h^3 v^3(5Z_1-2Z_3) \right. \nonumber \\
&&\left. + 2h^4 v^2(16Z_1-7Z_3) + 2h^5 v(4Z_1-3Z_3) - h^6 Z_3 \right. \nonumber \\
&&\left. - 4\left(v^4 + 3h v^3 + 4h^2 v^2 + 3h^3 v +h^4 \right)\la V_\mu V^\mu \ra \right. \nonumber \\
&&\left. + 2\left(v^4 + 4h v^3 + 6h^2 v^2 + 4h^3 v + h^4\right) \la V_\mu \sigma_3 \ra \la V^\mu \sigma_3 \ra \right\}~\label{Lxi2}
\eea
where we introduce $V_\mu\equiv U^\dagger D_\mu U$ as a shorter notation. It is to be observed that at $\mathcal{O}(\xi^3)$ order chiral dimension four operators also appear. Due to the verbosity, we list the HEFT operators in $\mathcal{L}(\xi^3)$ and their corresponding Wilson coefficients~\footnote{Here we use ``Wilson coefficient'' to denote the factor of each HEFT operators, which is a generic polynomial function of $h$. Such an abused notation should not cause confusion in the context.} in Tab.~\ref{tab:opsp2} and Tab.~\ref{tab:opsp4} at order $p^2$ and $p^4$ separately. 
\begin{table}[htbp]
\centering
\begin{tabular}{l l}
\hline\hline
Operators & $P(h)/\left[\xi^3/(Y_3 v^3) \right]/(4Z_1-Z_3)$ \\
\hline \hline
   $\la V_\mu V^\mu \ra $ & $-2h v^5  - 10 h^2 v^4 -22 h^3 v^3 $  
 $-23 h^4 v^2 -11 h^5 v -2 h^6 $\\
   \hline
   $\la V_\mu \s_ 3 \ra \la V^\mu \s_3\ra$ & $2h v^5  + 9h^2 v^4 
    +16 h^3 v^3+ 14 h^4 v^2+ 6 h^5 v+ h^6$   
   \\ \hline
   $D_\mu h D^\mu h$ & $20h^2 v^2  + 24h^3 v + 8 h^4$ \\ \hline\hline
$V(h)$ & \makecell[l]{$h^2 v^6 (4 Z_1 - Z_3)$+ $h^3 v^5 (16 Z_1 - 5 Z_3)+ \frac{1}{4} h^4 v^4 (60 Z_1 - 41 Z_3)$ \\ $-h^5 v^3 (4 Z_1 + 11 Z_3-\frac{1}{2} h^6 v^2 (24 Z_1 + 13 Z_3)-2 h^7 v (3 Z_1 + Z_3)$ \\ $-\frac{1}{4} h^8 (4 Z_1 + Z_3)+(-4 h^2 v^4 -10 h^3 v^3 -8 h^4 v^2 -2 h^5 v) Y_3^2/(4Z_1-Z_3)$}\\
 \hline
\end{tabular}
\caption{The $p^2$ operators and their corresponding Wilson coefficients at $\mathcal{O}(\xi^3)$ after the heavy fields $K$ and $\phi^\pm$ (equivalently $\phi_{1, 2}$) are integrated out. In the second column, the Wilson coefficient of each operator---a polynomial function of $h$---is normalized by a factor of $\xi^3/(Y_3 v^3)/(4Z_1-Z_3)$. The last row shows pure $h$ self-interactions.
~\label{tab:opsp2}}
\end{table}
\begin{table}[htbp]
    \centering
    \begin{tabular}{l l}
\hline   
\hline
Operator & $P(h)/\left[\xi^3/(Y_3 v^3)\right]$ \\
\hline\hline
$\la V_\mu V^\mu \ra \la V_\nu V^\nu \ra$ & $(h+v)^4$ \\
$\la V_\mu \s_3\ra \la V^\mu \s_3\ra \la V_\nu V^\nu \ra$ & $-2(h+v)^4$ \\
$\la V_\mu \s_3 \ra \la V_\nu \s_3 \ra \la V^\mu V^\nu \ra$ & $2(h+v)^4$ \\ \hline
$\la V_\mu V_\nu \s_3 \ra \la V^\mu \s_3 \ra D^\nu h$ & $-4(h+v)^3$ \\
$\la V_\mu V_\nu \s_3 \ra \la V^\nu \s_3 \ra D^\mu h$ & $4(h+v)^3$ \\ \hline
$\la V_\mu V^\mu \ra D_\nu h D^\nu h$ & $4(h+v)^2$ \\
$\la V_\mu \s_3\ra \la V^\mu \s_3 \ra D_\nu h D^\nu h$ & $-4(h+v)^2$ \\
$\la V_\mu V_\nu \ra D^\mu h D^\nu h$ & $-8(h+v)^2$ \\
$\la V_\mu \s_3 \ra \la V_\nu \s_3 \ra D^\mu h D^\nu h$ & $4(h+v)^2$ \\ \hline
$D_\mu h D^\mu h D_\nu h D^\nu h$ & $4$ \\ \hline
\end{tabular}
\caption{The $p^4$ operators and their corresponding Wilson coefficients at $\mathcal{O}(\xi^3)$. The Wilson coefficient of each operator is normalized by a factor $\xi^3/(Y_3 v^3)$.~\label{tab:opsp4}}
\end{table}

Note that the VEV $v$ shown above is the VEV of the doublet field. After rescaling it to the EW VEV $\vev$ (see Eq.~\eqref{vEW}), we can rewrite the effective Lagrangian into the standard HEFT LO form (Eq.~\ref{eq:HEFT}) with the coefficients listed in Tab.~\ref{tab:HEFT_couplings} in terms of ($Z_1, Z_2, Z_3, Y_3, \vev, \xi$) and Tab.~\ref{tab:HEFT_couplings_mh} in terms of ($m_h, Z_2, Z_3, Y_3, \vev, \xi$). For the operators of $\mathcal{O}(p^4)$, the complete results can be obtained similarly while in the following we are only interested in the $3$- or $4$-point interactions, therefore we only have to keep the terms without field $h$ in $P(h)$ and replace $v$ with $\vev$ in Tab.~\ref{tab:opsp4}.
\begin{table}[ht]
\centering
\begin{tabular}{c|cc}
\hline\hline
$\mathcal{O}(\xi)$ & $\xi^2$ & $\xi^3$ \\
\hline\hline
$c^k_1$ & $0$ & $0$ \\
$c^k_2$ & $4$ & $40(4Z_1-Z_3)\frac{\vev}{Y_3}$ \\
$c^k_3$ & $0$ & $48(4Z_1-Z_3)\frac{\vev}{Y_3}$ \\ \hline
$\Delta\kappa_3$ & $-2\left(2-\frac{Z_3}{Z_1}\right)$ & $\frac{2Z_3}{Z_1^2}\frac{Y_3}{\vev}-\frac{4(12Z_1^2-7Z_1 Z_3+Z_3^2)}{Z_1}\frac{\vev}{Y_3}$ \\
$\Delta\kappa_4$ & $-12\left(2-\frac{Z_3}{Z_1}\right)$ & $\frac{12Z_3}{Z_1^2}\frac{Y_3}{\vev}-\frac{8(4Z_1-Z_3)(7Z_1-5Z_3)}{Z_1}\frac{\vev}{Y_3}$ \\ \hline
$\Delta a$ & $4$ & $4(4Z_1-Z3)\frac{\vev}{Y_3}$ \\
$\Delta b$ & $16$ & $40(4Z1-Z3)\frac{\vev}{Y_3}$ \\ \hline
$\Delta\alpha$ & $2$ & $0$ \\
$\Delta a^\slashed{C}$ & $8$ & $8(4Z_1-Z_3)\frac{\vev}{Y_3}$ \\
$\Delta b^\slashed{C}$ & $12$ & $36(4Z_1-Z_3)\frac{\vev}{Y_3}$ \\ \hline\hline
$m_h^2$ & \multicolumn{2}{c}{$2\vev^2 Z_1 - 2\vev Y_3 \xi - 4v^2(4Z_1-Z_3)\xi^2 - 2\left[(4Z_1-Z_3)^2\vev^2-6Y_3^2\right]\frac{\vev}{Y_3}\xi^3$} \\ \hline
\end{tabular}
\caption{The Wilson coefficients of the HEFT $\mathcal{O}(p^2)$ operators and the squared mass of the Higgs $m_h^2$ for the triplet model in our redundant HEFT basis in terms of ($Z_1, Z_2, Z_3, Y_3, \vev, \xi$). All the couplings are shown up to $\mathcal{O}(\xi^3)$.~\label{tab:HEFT_couplings}}
\end{table}

\begin{table}[h!]
\centering
\begin{tabular}{c|cc}
\hline\hline
$\mathcal{O}(\xi)$ & $\xi^2$ & $\xi^3$ \\
\hline\hline
$c^k_1$ & $0$ & $0$ \\
$c^k_2$ & $4$ & $40\left(2\frac{m_h^2}{\vev^2}-Z_3\right)\frac{\vev}{Y_3}$ \\
$c^k_3$ & $0$ & $48(4Z_1-Z_3)\frac{\vev}{Y_3}$ \\ \hline
$\Delta\kappa_3$ & $-4\left(1-\frac{\vev^2}{m_h^2}Z_3\right)$ & $-4\frac{6m_h^4-7m_h^2\vev^2 Z_3+2\vev^4 Z_3^2}{m_h^2\vev Y_3}$ \\
$\Delta\kappa_4$ & $-24\left(1-\frac{\vev^2}{m_h^2}Z_3\right)$ & $-8\frac{(7m_h^2-10\vev^2 Z_3)(2m_h^2-\vev^2 Z_3)}{m_h^2\vev Y_3}$ \\ \hline
$\Delta a$ & $4$ & $4\left(\frac{2m_h^2}{\vev^2}-Z3\right)\frac{\vev}{Y_3}$ \\
$\Delta b$ & $16$ & $40\left(\frac{2m_h^2}{\vev^2}-Z3\right)\frac{\vev}{Y_3}$ \\ \hline
$\Delta\alpha$ & $2$ & $0$ \\
$\Delta a^\slashed{C}$ & $8$ & $8\left(\frac{2m_h^2}{\vev^2}-Z3\right)\frac{\vev}{Y_3}$ \\
$\Delta b^\slashed{C}$ & $12$ & $36\left(\frac{2m_h^2}{\vev^2}-Z3\right)\frac{\vev}{Y_3}$ \\ \hline
\end{tabular}
\caption{The Wilson coefficients of the HEFT $\mathcal{O}(p^2)$ operators and the squared mass of the Higgs $m_h^2$ for the triplet model in our redundant HEFT basis in terms of ($m_h, Z_2, Z_3, Y_3, \vev, \xi$). All the couplings are shown up to $\mathcal{O}(\xi^3)$.~\label{tab:HEFT_couplings_mh}}
\end{table}

As we mentioned at the beginning of this section, here we perform the matching by solving the EoMs of the heavy states. One can do this matching of predictions at the amplitude level. To have a cross-check of our results, we have also performed a matching at the amplitude level of some specific processes with up to four external particles, and we find that the two methods arrive at the same results. However, it is difficult to obtain the complete EFT truncated at a certain order by using the amplitude method without knowing the resulting operators in advance~\footnote{Here we treat different terms in a Wilson coefficient polynomial as different operators since they contribute to different $n$-point amplitudes. More precisely, the overcomplete operators can be determined by dimensional analysis, but this generally results in a rather large set which can generate amplitudes with too many external particles. For example, to determine the coefficient function for $\braket{V_\mu V^\mu}\braket{V_\nu V^\nu}$, we can assume that the function has the form of $c_1 v^4+c_2 v^3 h+c_3 v^2 h^2+c_4 v h^3+c_5 h^4$ through dimensional analysis, and then solve these coefficients $c_i$ order by order by calculating up to 8-point amplitudes. }, while the functional method can deal with the coefficient polynomial as a single object, significantly simplifying the matching process.

\subsection{One-loop level}
From the perspectives for testing the effects of the BSM physics from the Higgs properties, it is also important to consider the local operators describing $h\rightarrow \gamma\gamma$ and $h\rightarrow\gamma Z$ transitions, which are loop suppressed in the SM. The EFT corrections are then at the 1-loop order as the leading contributions. In the HEFT, the relevant operators show up at the next to leading order with chiral dimension four, which are given as follows
\begin{align}
    \mathcal{L}_\text{HEFT}\supset -c_{hBB}\frac{h}{v}\text{Tr}[\hat{B}_{\mu\nu}\hat{B}^{\mu\nu}]-c_{hWW}\frac{h}{v}\text{Tr}[\hat{W}_{\mu\nu}\hat{W}^{\mu\nu}]+c_{hBW}\frac{h}{v}\text{Tr}[U\hat{B}_{\mu\nu}U^\dagger\hat{W}^{\mu\nu}]
\end{align}
where for simplicity we introduce the hat notation
\bea
\hat{B}_{\mu\nu}&=&D_\mu\hat{B}_\nu-D_\nu\hat{B}_\mu,\, \hat{B}_\mu= g^\prime B_\mu \sigma_3/2 , \\
\hat{W}_{\mu\nu}&=&D_\mu\hat{W}_\nu-D_\nu\hat{W}_\mu+i[\hat{W}_\mu, \hat{W}_\nu],\,
\hat{W}_\mu= g W_{i \mu } \sigma_i/2, 
\eea
which generate the effective operators
\begin{align}
    -\frac{1}{2}\frac{e^2}{v}c_{h\gamma\gamma}h A_{\mu\nu}A^{\mu\nu}-\frac{egc_{\theta_w}}{v}c_{h\gamma Z}h A_{\mu\nu}Z^{\mu\nu}
\end{align}
where
\begin{align}
    c_{h\gamma\gamma}&=c_{hBB}+c_{hWW}-c_{hBW}, \\
    c_{h\gamma Z}&=\frac{1}{c_{\theta_w}^2}\left(-c_{hBB}s_{\theta_w}^2+c_{hWW}c_{\theta_w}^2-\frac{1}{2}c_{hBW}(c_{\theta_w}^2-s_{\theta_w}^2)\right)
\end{align}
Though to match to the HEFT one should obtain the coefficients $C_{HBB, HWW, HBW}$, the dominant effects from the one loop matching only present in the decay processes $h\rightarrow \gamma\gamma$ and $h\rightarrow\gamma Z$. Therefore here we only give the coefficients $c_{H\gamma\gamma, H\gamma Z}$ and leave the complete 1-loop results for future work. With a direct calculation, we find that
\begin{align}
    c_{h\gamma\gamma}&=\frac{Z_3}{48\pi^2}\frac{\vev}{Y_3}\xi+\frac{Z_3 m_h^2+15Y_3^2}{180\pi^2}\frac{1}{Y_3^2}\xi^2+\cdots \\
    c_{h\gamma Z}&=\frac{(2c_{\theta_w}^2-1)Z_3}{96c_{\theta_w}^2\pi^2}\frac{\vev}{Y_3}\xi+(2c_{\theta_w}^2-1)\frac{8c_{\theta_w}^2(Z_3 m_h^2+15Y_3^2)+Z_3 g^2\vev^2}{2880c_{\theta_w}^4\pi^2}\frac{1}{Y_3^2}\xi^2+\cdots
\end{align}
Unlike the Wilson coefficients obtained from the tree level matching, these coefficients show up at order $\mathcal{O}(\xi)$.

\section{Phenomenology~\label{sec:mat2SMEFT}}
In this section, we perform a numerical study on the tree-level scattering processes $hh\rightarrow hh$, $WW\rightarrow hh$. We also consider the process $ZZ\rightarrow hh$ since the custodial symmetry violating term generated at the leading order in the HEFT, which is not the case in other well-studied models like the scalar singlet extension of the SM and the 2HDM.
 
It is believed that SMEFT is less general than HEFT, but it is more commonly used and well studied in the BSM physics. So we use SMEFT as a reference EFT. Through comparing HEFT's results with its, we could know the quality of matching. Usually, one expects there are some minor differences between HEFT and SMEFT due to their different power counting in the above processes since the longitudinal mode scattering and Higgs scattering provides important probe of electroweak symmetry breaking. Furthermore, for some parameter regions where SMEFT breaks down, HEFT is still available. It is also interesting to show that our matching results can reveal this point explicitly. Therefore we first perform the SMEFT matching to th RHTE, and then we compare its phenomenological results with HEFT's and study their convergence properties. It is a good verification for our HEFT matching procedure (the strategy of power counting). Finally we show how HEFT accommodates larger parameter spaces than SMEFT. 

\subsection{Matching to the SMEFT}
The matching to the SMEFT must be done by integrating out the heavy degrees of freedom of the triplet model in the unbroken phase. Yet the mass states are not well-defined before spontaneous symmetry breaking. Further the VEV of the triplet $v'$ (equivalent $\xi$) also becomes meaningless in this case. The question then arises as to how can the matching be performed (the heavy states be integrated out).

The trouble caused by the definition of the triplet vev $v'$ is never a real problem since it is originated from the set of independent parameters of the UV model that we chose. The freedom of the choices for independent parameters allows one to change between different parameter sets without affecting the underlying UV theory. Therefore a certain value of $\xi$ in our parameter set can always be convert to some other values of parameters in another set with meaningful definition in the unbroken phase. However in the EFT, things become complicated due to the finite truncation of power counting parameters and the different scaling behaviors in different parameter set choices. The scaling behavior dependence of the EFT matching has already been noticed and discussed in Ref.~\cite{Dawson:2023oce}. We also notice that generally the conversion functions are not polynomials with the highest power being consistent with the designed parameter power of the EFT. Therefore further expansion and truncation should be carried out for the conversion functions, which indicates that the EFT expansions are not commutable with different power counting methods. A more serious problem could arise if the conversion functions are singular at the expansion parameter of one of the EFTs, which without a doubt can demolish the power counting of that EFT. There are no differences for different choices of parameters in the full theory, but they indeed parameterize different subspaces of the parameter space of the UV theory in different EFTs, which though can be rather close to each other, and could have significant consequences in the predictions.

The answer to the first problem has to do with the decoupling limit ($Y_2\rightarrow\infty$). It is easy to check that as long as $Y_3$ has a fixed value the non-SM neutral Higgs and charged Higgs become mass states and degenerate, which can be embedded into a gauge triplet in this decoupling scenario. Therefore, the triplet $\Sigma$ can be integrated out as a whole and the resulting Lagrangian will be an expansion in inverse powers of $Y_2$. This parameter thus can be identified as a power counting parameter and the resulting EFT is in the format of the SMEFT.

The tree-level SMEFT matching of the triplet model has been calculated by several groups. The dimension-6 results have been computed in Refs~\cite{Henning:2014wua, Corbett:2017ieo, deBlas:2017xtg}, and the results to dimension-8 have been calculated in Refs~\cite{Corbett:2021eux, Ellis:2023zim}. We perform this exercise again using the Covariant Derivative Expansion (CDE) method but ignoring the fermion sector. The Wilson coefficients of operators involving fermions are always suppressed by tiny Yukawa couplings unless one wants to consider some phenomenology related to top physics. Therefore we can safely ignore the fermion sector in the following discussion. Our result is consistent with Refs.~\cite{Corbett:2021eux, Ellis:2023zim} and listed below in Tab.~\ref{tab:SMEFT_couplings}.
\begin{table}[h!]
\begin{center}
\begin{tabular}{c|c}
\hline
\multicolumn{2}{c}{$dim$-4} \\
\hline\hline
$C_{H^4}$ & $-Z_1+\frac{Y_3^2}{2Y_2^2}+\frac{2Y_1^2 Y_3^2}{Y_2^4}+\frac{6Y_1^4 Y_3^2}{Y_2^6}$ \\
\hline
\multicolumn{2}{c}{$dim$-6} \\
\hline\hline
$C_H$ & $\frac{Y_3^2}{2Y_2^4}\left[\left(8Z_1-Z_3\right)\left(1+\frac{4Y_3^2}{Y_2^2}\right)-\frac{2Y_3^2}{Y_2^2}\left(2+\frac{11Y_1^2}{Y_2^2}+\frac{6Y_1^4}{Y_2^4}+\frac{9Y_1^6}{Y_2^6}\right)\right]$ \\
$C_{HD}$ & $-\frac{2Y_3^2}{Y_2^4}\left(1+\frac{4Y_1^2}{Y_2^2}\right)$ \\
$C_{H\Box}$ & $\frac{Y_3^2}{2Y_2^4}\left(1+\frac{4Y_1^2}{Y_2^2}\right)$ \\
\hline
\multicolumn{2}{c}{$dim$-8} \\ 
\hline\hline
$C_{H^8}$ & \makecell[c]{$\frac{Y_3^2}{4Y_2^6}\left\{2\left(4Z_1-Z_3\right)^2-\frac{Y_3^2}{Y_2^2}\left(72Z_1+Z_2-20Z_3\right)+\frac{4Y_3^4}{Y_2^4}\left[7+\frac{Y_1^2}{Y_3^2}\right.\right.$ \\ $\left.\left.\left(-40Z_1+11Z_3\right)+\frac{2Y_1^2}{Y_2^2}\left(11+\frac{3Y_1^2}{Y_3^2}\left(-11Z_1+Z_3\right)\right)\right]+\frac{39Y_1^4}{Y_2^4}\right\}$} \\
$C_{H^6}^{(1)}$ & $-\frac{Y_3^4}{Y_2^8}$ \\     
$C_{H^6}^{(2)}$ & $\frac{2Y_3^2}{Y_2^6}\left(-4Z_1+Z_3+\frac{4Y_3^2}{Y_2^2}\right)$ \\     
$C_{H^4}^{(1)}$ & $\frac{4Y_3^2}{Y_2^6}$ \\     
$C_{H^4}^{(3)}$ & $-\frac{2Y_3^2}{Y_2^6}$ \\  
\hline\hline
\end{tabular}
\caption{Dimension-6 and -8 Wilson coefficients of the bosonic operators resulting from the tree-level matching of the RHTE to the SMEFT. $C_{H^4}$ represents the coefficient of the Higgs doublet quartic interaction after matching.~\label{tab:SMEFT_couplings}}
\end{center}
\end{table}

Our SMEFT matching results are shown in a parameter set $(Z_1, Z_2, Z_3,Y_1, Y_2, Y_3 )$ without any physical meanings as in all other SMEFT matching references. But as we discussed in Sec.~\ref{sec:mat2HEFT}, we use ($m_h, Z_2, Z_3, Y_3, v, \xi$) as our input parameters. Therefore, as mentioned previously, a certain conversion based on Eq.~\eqref{Y1Y2} and \eqref{eq:NSmass2} 
is performed implicitly when we present the plots. One can explicitly see that these two parameter choices generally are preferred by two different UV parameter regions and the transition of the SMEFT failure is also shown explicitly.

One should note that there is another dimensionful parameter $Y_3$ in the RHTE and we say nothing about it during the SMEFT matching. One would expect that it does not cause any problems until $Y_3$ is larger than $Y_2$. However once $Y_3$ becomes the largest dimensionful parameter, treating the term $2Y_3 H^\dagger\Sigma H$ as a perturbation of the free scalar fields seems questionable, and so does the above SMEFT results. However in our choice of the parameter set, we always assume the value of parameter $\xi$ is small, the expansion of $\xi$ in the HEFT matching seems still valid. Indeed in the next section we will show that the SMEFT loses its predictive power when $Y_3$ becomes larger than $Y_2$. Since the purpose of this paper is to illustrate the results of our HEFT matching, we will not further explore this scenario~\footnote{Up to our knowledge, there is no consistent method to obtain a SMEFT in this case, though people never point out this explicitly. Even when one directly tries to calculate the physics observables in the UV model, the correct way in quantum field theory might still be unclear. One can compare this scenario to the classical coupled harmonic oscillators, in which collective modes are more suitable degrees of freedom than normal mode in the strongly coupled region.}.

\subsection{Numerical results~\label{sec:NumRSLT}}
The results that follow are obtained via FeynArts~\cite{Hahn:2000kx} and FormCalc~\cite{Hahn:1998yk}, as well as FeynRules~\cite{Alloul:2013bka} and SmeftFR~\cite{Dedes:2017zog, Dedes:2019uzs, Dedes:2023zws} for generating Feynman rules in different models. The results shown are in the range allowed by the theoretical constraints but we do not check whether they survive under the direct neutral and charged Higgs searches. However, the masses of the extra scalars in the region of parameter space we considered are heavier than $\mathcal{O}(500)$ GeV, which should be safe from the experimental detection. Further we only consider values of $\xi$ less than 0.02, equivalent to the triplet VEV less than $4.9$ GeV~\cite{ParticleDataGroup:2024cfk}, which does not conflict with the electroweak precision tests in a large portion of the parameter space.
\begin{figure}[htb!]
\centering
\includegraphics[width=0.48\textwidth]{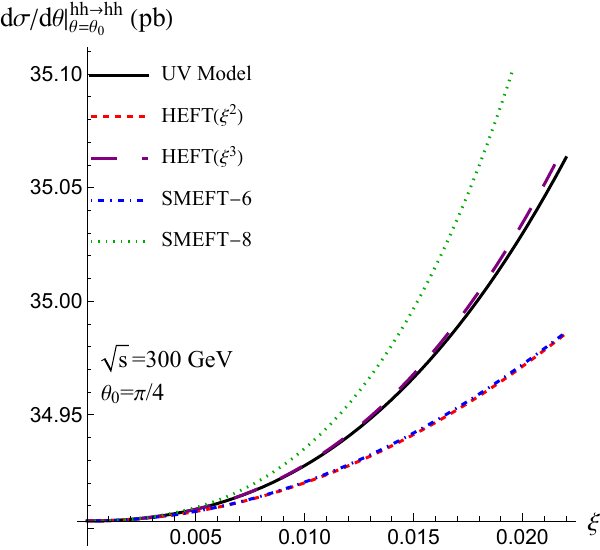}
\includegraphics[width=0.48\textwidth]{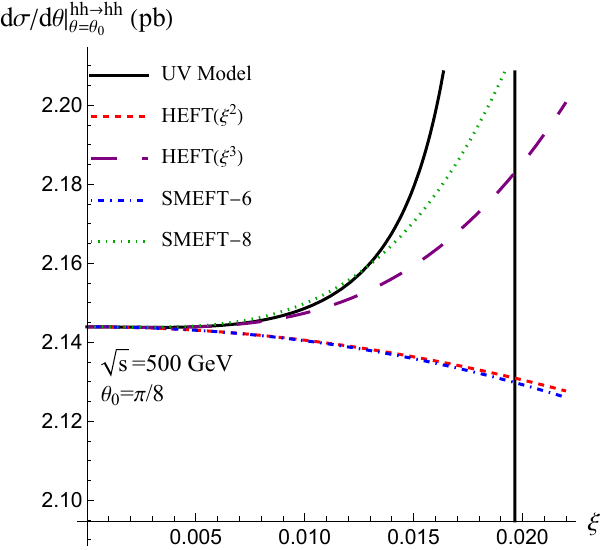}
\caption{Comparison between the UV model and the HEFT, the SMEFT $dim$-6 (SMEFT-6) and the SMEFT $dim$-8 (SMEFT-8) approaches to it in the differential cross-section of $hh\rightarrow hh$, for a center-of-mass energy $\sqrt{s}$ and a scattering angle $\theta_0$. On both panels, we take $Y_3=39.75$ GeV, $Z_2=1$ and $Z_3=0.759$, while all other parameters can be fixed by the SM inputs for a certain value of $\xi$.} 
\label{fig:hh2hh_LY3}
\end{figure}

In Fig.~\ref{fig:hh2hh_LY3}, we present the differential cross section $hh\rightarrow hh$, for two different values of the total collision energy: $\sqrt{s}=300$ GeV (left panel) and $\sqrt{s}=500$ GeV (right panel), with a low value of $Y_3$ ($Y_3=39.75$ GeV). In the left panel, we find that involving the effects of the $dim$-8 operators could not reduce the absolute difference between the SMEFT and the full model. Although the HEFT provides a comparable replication to the dimension 6 SMEFT at the leading order of $\mathcal{O}(\xi)$, the $\mathcal{O}(\xi^3)$ truncation of the HEFT successfully reproduces the UV model. The failure of the SMEFT can easily be understood by noticing that the masses of the BSM states can be as low as $495$ GeV, or the equivalent value of $Y_2\simeq 470$ GeV , when $\xi$ approaches $0.02$, indicating that the SMEFT converges slowly to the full model. This is clearer in the right panel of Fig.~\ref{fig:hh2hh_LY3}. The divergence of the UV model (the black vertical line) at around $\xi=0.0196$ is due to the presence of heavy scalars with masses below the center-of-mass energy. Both the SMEFT and the HEFT lose the predictive power there. Away from such a value of $\xi$, both EFTs give similar approximations at the same order and the second order, SMEFT-8 and HEFT($\xi^3$), is a very good replication of the RHTE.
\begin{figure}[htb!]
\centering
\includegraphics[width=0.48\textwidth]{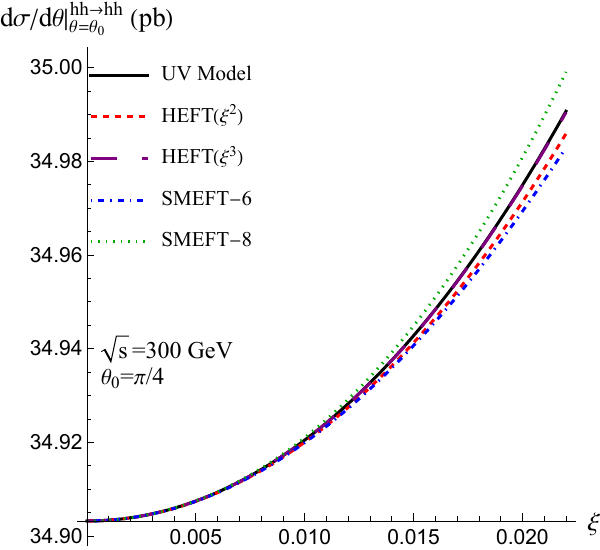}
\includegraphics[width=0.48\textwidth]{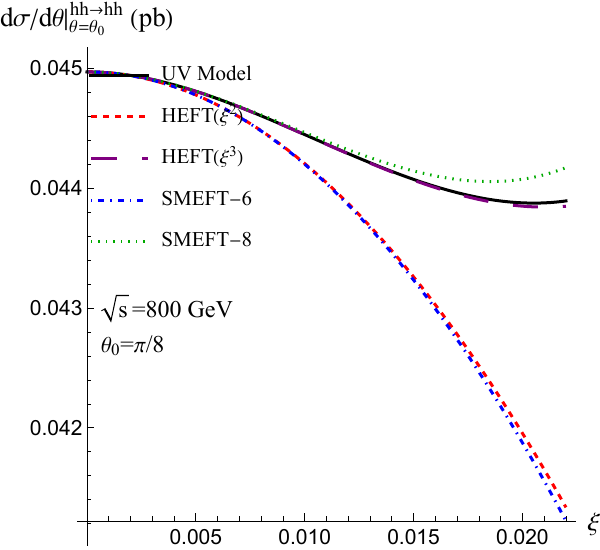}
\caption{Comparison between the UV model and the HEFT, the SMEFT $dim$-6 (SMEFT-6) and the SMEFT $dim$-8 (SMEFT-8) approaches to it in the differential cross-section of $hh\rightarrow hh$, for a center-of-mass energy $\sqrt{s}$ and a scattering angle $\theta_0$. On both panels, we take $Y_3=730.14$ GeV, $Z_2=1$ and $Z_3=0.758$, while all other parameters can be fixed by the SM inputs for a certain value of $\xi$.} 
\label{fig:hh2hh_HY3}
\end{figure}

We present similar plots but for a large value of $Y_3$ ($Y_3=730.14$ GeV) for different values of the center-of-mass energy ($\sqrt{s}=300$ GeV for the left panel and $\sqrt{s}=800$ GeV for the right panel) in Fig.~\ref{fig:hh2hh_HY3}. In this case, both EFTs are still good replications of the RHTE while the HEFT $\mathcal{O}(\xi^3)$ shows a better description for larger values of $\xi$. The improvement of the SMEFT can be understood by noticing that the lowest value of $Y_2$ is around $2115$ GeV, which is far above both the electroweak scale and collision energy, and makes the SMEFT expansion under control.
\begin{figure}[htb!]
\centering
\includegraphics[width=0.48\textwidth]{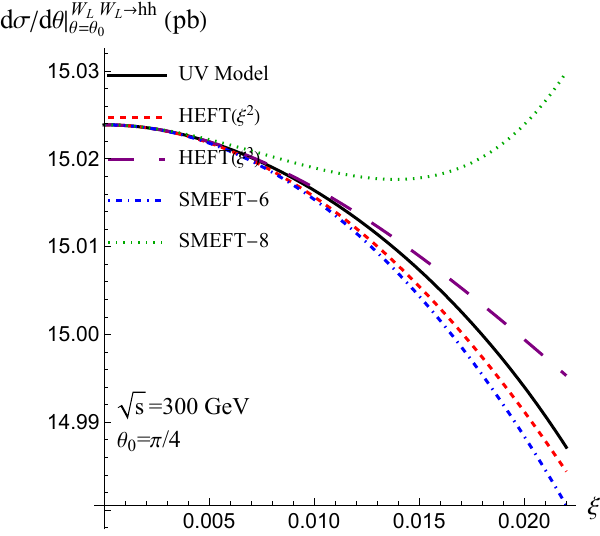}
\includegraphics[width=0.48\textwidth]{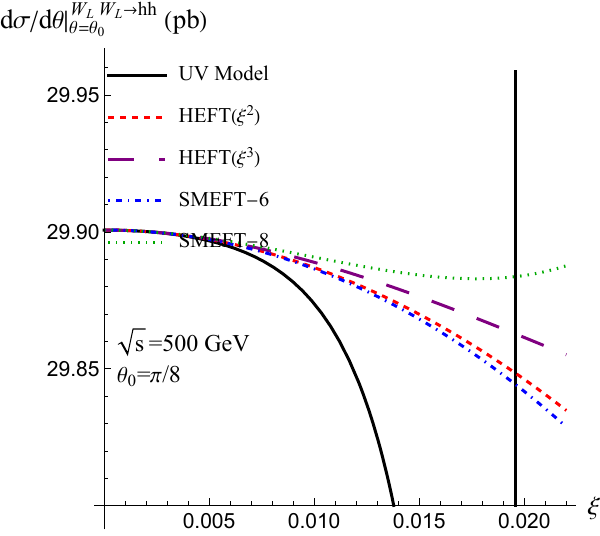}
\includegraphics[width=0.48\textwidth]{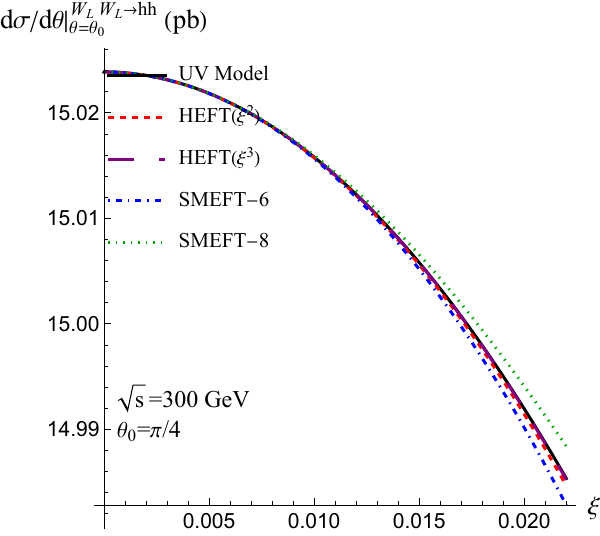}
\includegraphics[width=0.48\textwidth]{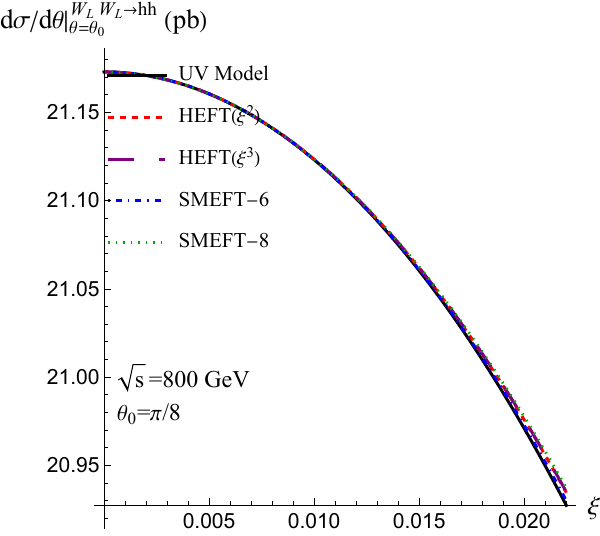}
\caption{Comparison between the UV model and the HEFT, the SMEFT $dim$-6 and the SMEFT $dim$-8 approaches to it in the differential cross-section of $W_L W_L\rightarrow hh$, for different center-of-mass energy $\sqrt{s}$'s and scattering angle $\theta_0$'s. For the top panels, we assume $Y_3=39.75$ GeV, $Z_2=1$ and $Z_3=0.759$, while we choose $Y_3=730.14$ GeV, $Z_2=1$ and $Z_3=0.758$ in the bottom panels.} 
\label{fig:ww2hh}
\end{figure}
\begin{figure}[htb!]
\centering
\includegraphics[width=0.48\textwidth]{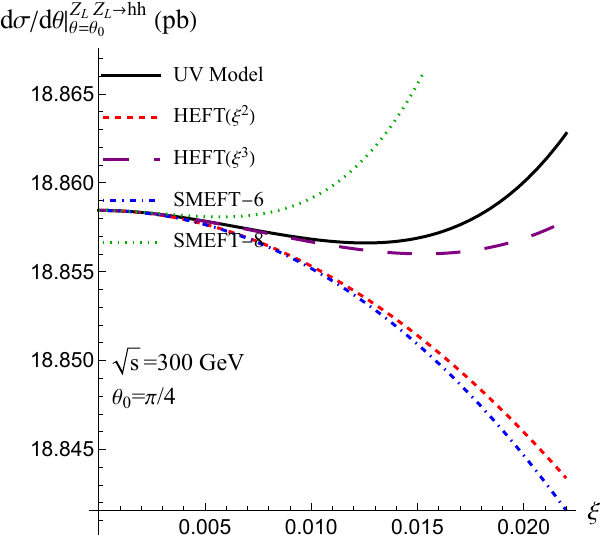}
\includegraphics[width=0.48\textwidth]{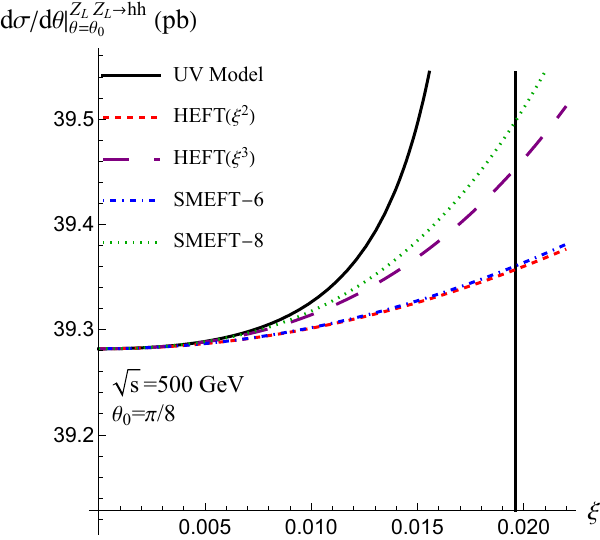}
\includegraphics[width=0.48\textwidth]{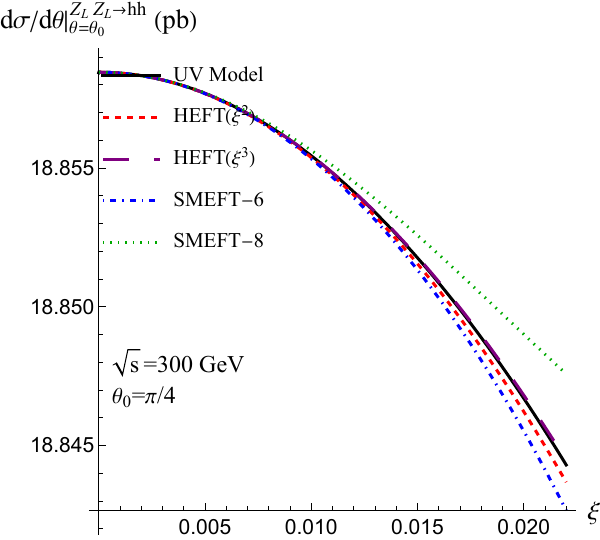}
\includegraphics[width=0.48\textwidth]{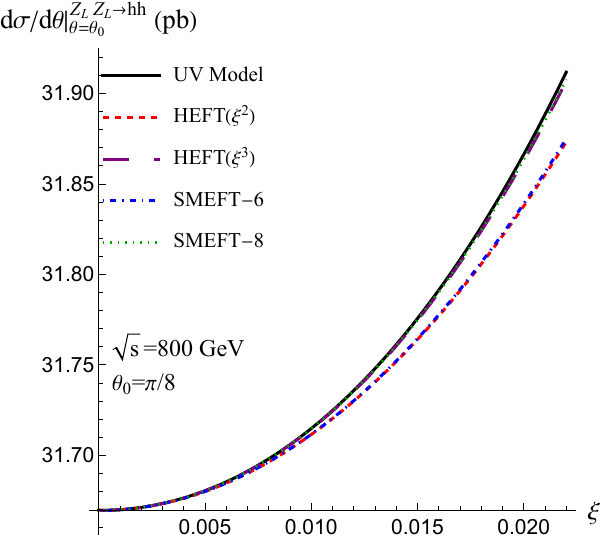}
\caption{Comparison between the UV model and the HEFT, the SMEFT $dim$-6 and the SMEFT $dim$-8 approaches to it in the differential cross-section of $Z_L Z_L\rightarrow hh$, for different center-of-mass energy $\sqrt{s}$'s and scattering angle $\theta_0$'s. The parameters are chosen similar to Fig.~\ref{fig:zz2hh}.} 
\label{fig:zz2hh}
\end{figure}

We now discuss the scattering processes of $WW\rightarrow hh$ and $ZZ\rightarrow hh$. The similar plots are shown in Fig.~\ref{fig:ww2hh} and Fig.~\ref{fig:zz2hh}. We only consider the longitudinal-mode scattering since 1) the contributions from transverse modes can be neglected and 2) at the high energy limit the longitudinal modes are just Goldstones required by the Goldstone equivalence theorem, which are deeply related to the electroweak symmetry breaking. Similar characteristics hold for $WW\rightarrow hh$ and $ZZ\rightarrow hh$. The $\mathcal{O}(\xi^3)$ corrections significantly improve the quality of the replication of the UV model in the HEFT and it provides the best approximation to reproduce the UV predictions for both large and small values of the collision energy. it is notable that the SMEFT yields a very poor replication when $Y_3$ is small (top panels).

\begin{figure}[h!]
\centering
\includegraphics[width=0.48\textwidth]{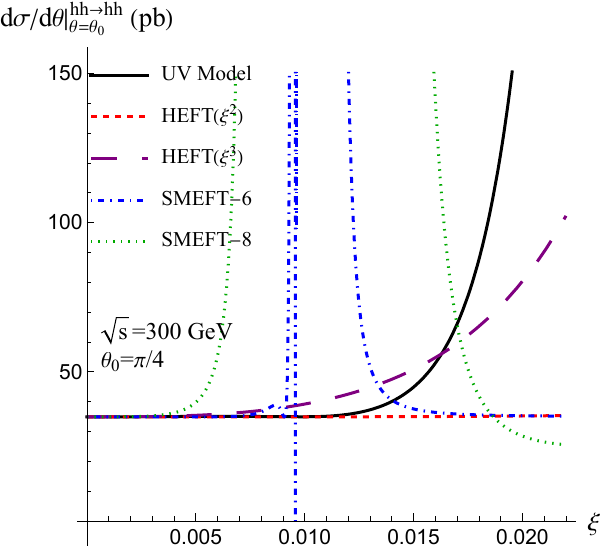}
\includegraphics[width=0.48\textwidth]{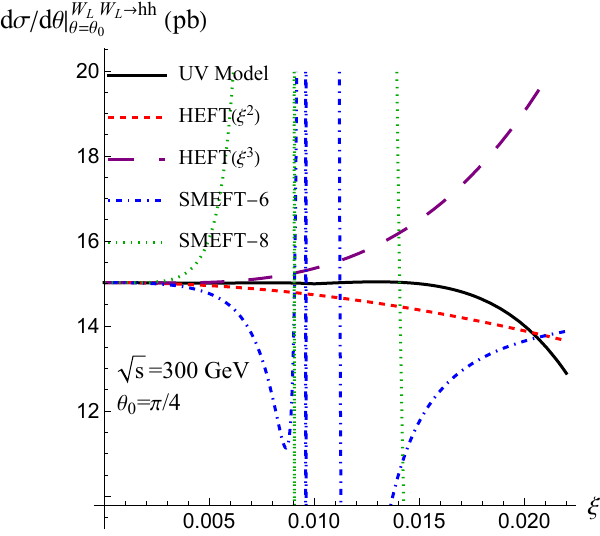}
\includegraphics[width=0.48\textwidth]{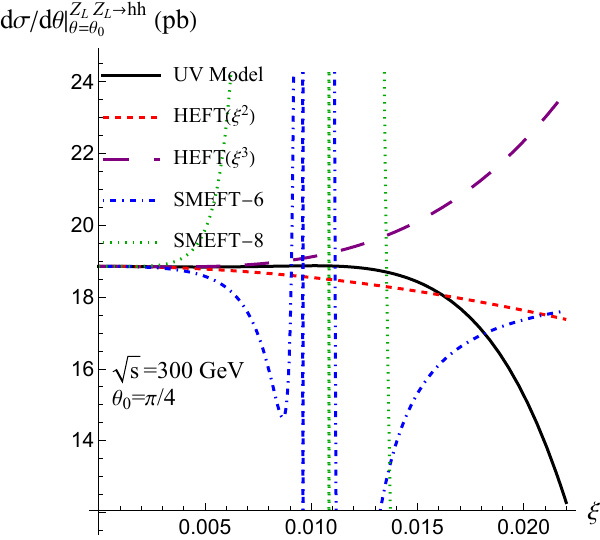}
\caption{The differential tree-level cross sections for $hh\rightarrow hh$ (top left panel), $WW\rightarrow hh$ (top right panel) and $ZZ\rightarrow hh$ (bottom panel) in the RHTE, in the HEFT at $\mathcal{O}(\xi^2)$ and $\mathcal{O}(\xi^3)$, and in the SMEFT at the $dim$-6 and $dim$-8 order for center-of-mass energy $\sqrt{s}=300$ GeV and scattering angle $\theta_0=\pi/4$. We take $Y_3=24.65$ GeV, $Z_2=1$ and $Z_3=10$.} 
\label{fig:NY2_s300}
\end{figure}

To understand this observation, we show in Fig.~\ref{fig:NY2_s300} ($\sqrt{s}=300$ GeV and $\theta_0=\pi/4$) for an extreme value of $Y_3$: $24.65$ GeV. To ensure that the resonances are heavy enough, we choose a large value of $Z_3$ ($Z_3=10$). All the channels constitute clear deviations from the UV results from the SMEFT predictions. With explicit calculations, one can see that $Y_2^2$ becomes smaller when $\xi$ becomes larger, and changes its sign (from positive to negative) at $\xi=0.01$. The smallness of $Y_2^2$ obversely indicates the failure of the standard SMEFT expansion in $1/Y_2^{2n}$. One should note that this is not just a numerical invalidation in the series expansion. A negative value of $Y_2^2$ means that even without the doublet, the triplet itself could develop a VEV, which breaks the electroweak symmetry $SU(2)_L\times U(1)_Y$. In fact a physical phase transition occurs at $\xi=0.01$. In such a case, the electroweak symmetry possessed by the SMEFT is not the correct symmetry. Without doubt, it cannot reproduce the full theory's results while the HEFT still works. This failure of the SMEFT has already been observed by the authors of Ref.~\cite{Cohen:2020xca}. However, we note that before $Y_2^2$ becomes negative, the value of $Y_3$ has already been larger than $Y_2$ at $\xi=9.98\times 10^{-3}$. As discussed previously, this may have already caused our SMEFT matching results to crash. Such an earlier losing prediction powers of the SMEFT can also be observed in the figures. We should emphasize that here we choose a large value of $Z_3$ not only to make the phase transition and the failure of the SMEFT clearer but also to ensure the BSM states are heavy enough (so the EFT description is valid), but the general conclusion that the HEFT provides a more accurate description than the SMEFT for small values of $Y_3$ holds for all the values of $Z_3$.

\begin{figure}[h!]
\centering
\includegraphics[width=0.48\textwidth]{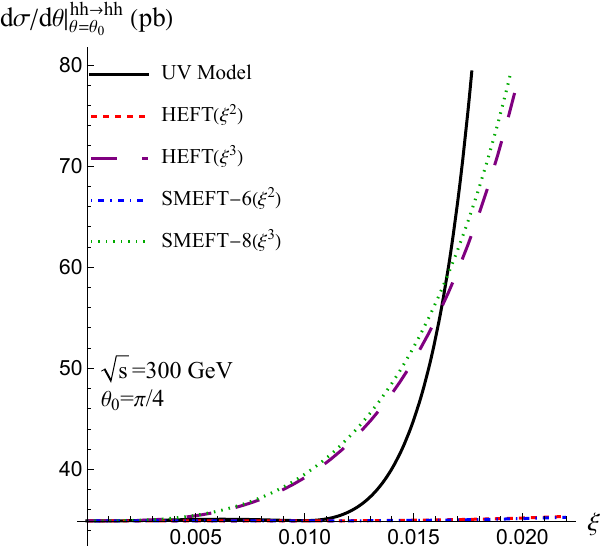}
\includegraphics[width=0.48\textwidth]{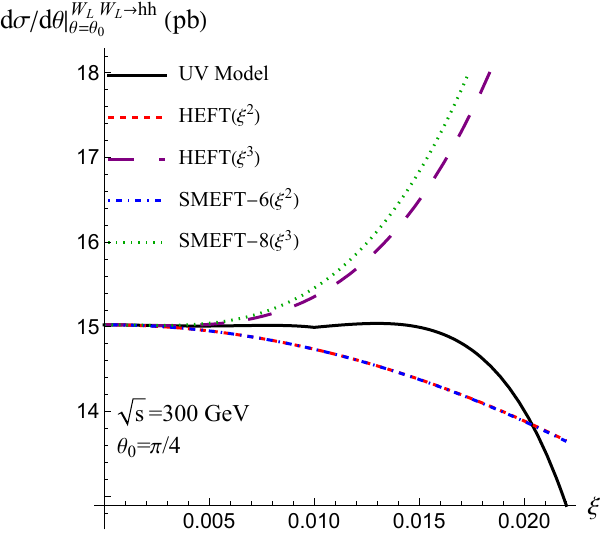}
\includegraphics[width=0.48\textwidth]{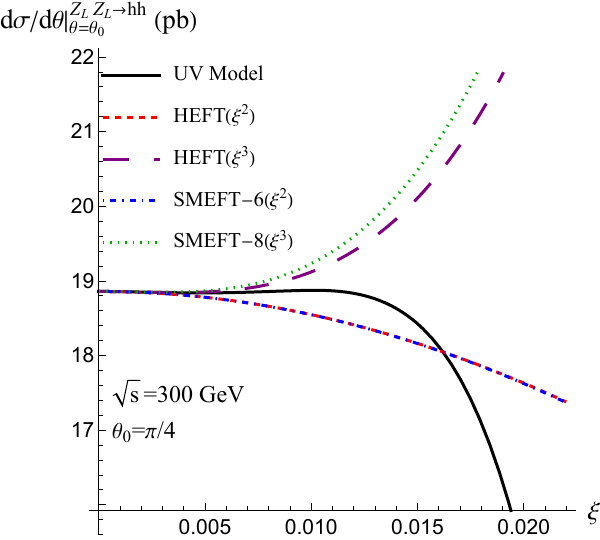}
\caption{The same as in Fig.~\ref{fig:NY2_s300}, but with the SMEFT coefficients truncated at a certain order of $\xi$. SMEFT-6($\xi^2$) and SMEFT-8($\xi^2$) represent that the corresponding SMEFT Wilson coefficients upto $dim$-6 and $dim$-8 operators are truncated at order $\mathcal{O}(\xi^2)$ and $\mathcal{O}(\xi^3)$ respectively.} 
\label{fig:NY2_s300_xi}
\end{figure}

We end this section by considering a ``consistent expansion'' in the different EFTs. With ``consistent expansion'', we mean that the same input parameters are used and the Wilson coefficients are expanded to the same order of the expansion parameters ($\xi$ in our case). This kind of expansion is also considered in Ref.~\cite{Dawson:2023ebe} for the 2HDM. To obtain a ``consistent'' SMEFT with the HEFT, following the method presented in Ref.~\cite{Dawson:2023ebe}, one can simply use the relationship between $Y_2$ and the input parameters $(m_h, Z_2, Z_3, Y_3,  v, \xi)$ to replace all $Y_2$'s in Tab.~\ref{tab:SMEFT_couplings}, and then truncate all coefficients to order $\mathcal{O}(\xi^3)$. The results obtained are listed in Tab.~\ref{tab:SMEFT_couplings_xi}. We can regenerate all the plots shown previously with the new coefficients. We only show plots similar to Fig.~\ref{fig:NY2_s300} with the exact same choice of parameter values in Fig.~\ref{fig:NY2_s300_xi}. We find that this time the SMEFT does not have a singular behavior and presents predictions comparable to those of the HEFT at $\mathcal{O}(\xi^3)$. Even though we do not show them explicitly, the SMEFT always shows improvements and becomes as good as the HEFT in other cases. One may naively consider that this consistency is not surprising. From the point of view of diagram approach of EFT matching, SMEFT and HEFT only provide different ways to parameterize the Lorentz and gauge structures in a $\xi$ expansion of the amplitude calculated in UV theory. The increasing difference between the SMEFT-8($\xi^3$) and HEFT($\xi^3$) at large values of $\xi$ is simply due to the fact that HEFT($\xi^3$) is exactly at order $\mathcal{O}(\xi^3)$ while SMEFT-8($\xi^3$) misses the contributions of higher dimension operators (such as $dim$-10 operator $(H^\dagger H)^5$). The standard canonical dimension in the SMEFT or the chiral dimension in the HEFT only provides a method to organize the operators but can be not the real expansion parameter in the matching procedure. Therefore, one might conclude that the SMEFT and HEFT matchings to the RHTE are identical to $\mathcal{O}(\xi^3)$ as in Ref.~\cite{Dawson:2023ebe}. However, we would like to point out that the above method to obtain the $\xi^3$ order SMEFT and the argument are incorrect. Note that the SMEFT Wilson coefficients in Tab.~\ref{tab:SMEFT_couplings} have a pole at $\xi=0.01$, which implies that the expansions of them at $\xi=0$ are only valid within a radius of convergence $|\xi|<0.01$. So the $\xi$ expansion fails beyond $\xi>0.01$, and the agreement between the SMETF-8($\xi^3$) and HEFT($\xi^3$) is superficial and might be just accidental coincidences. In our case, this is very different from the one considered in Ref.~\cite{Dawson:2023ebe}, where the SMEFT scale $\Lambda$ is chosen as an independent parameter and the decoupling limit is assumed. There is no question to presenting the UV amplitude expansion in terms of the SMEFT operators, while the correct Wilson coefficients should be calculated directly by the diagram method, which is beyond the scope of this work.
\begin{table}[h!]
\begin{center}
\begin{tabular}{c|c|c}
\hline
\multicolumn{2}{c}{$dim$-6} \\
\hline\hline
$C_H$ & $2\frac{4m_h^2-Z_3\vev^2}{\vev^4}\xi^2-4\frac{(4m_h^2-Z_3\vev^2)(2m_h^2-Z_3\vev^2)}{\vev^5 Y_3}\xi^3$ \\
$C_{HD}$ & $-8\frac{1}{\vev^2}\xi^2+16\frac{2m_h^2-Z_3\vev^2}{\vev^3 Y_3}\xi^3$ \\
$C_{H\Box}$ & $2\frac{1}{\vev^2}\xi^2-4\frac{2m_h^2-Z_3\vev^2}{\vev^3 Y_3}\xi^3$ \\
\hline
\multicolumn{2}{c}{$dim$-8} \\ 
\hline\hline
$C_{H^8}$ & $-4\frac{\left(2m_h^2-Z_3\vev^2\right)^2}{\vev^7 Y_3}\xi^3$ \\
$C_{H^6}^{(1)}$ & $0$ \\     
$C_{H^6}^{(2)}$ & $-16\frac{2m_h^2-Z_3\vev^2}{\vev^5 Y_3}\xi^3$ \\     
$C_{H^4}^{(1)}$ & $32\frac{1}{\vev^3 Y_3}\xi^3$ \\     
$C_{H^4}^{(3)}$ & $-16\frac{1}{\vev^3 Y_3}\xi^3$ \\  
\hline\hline
\end{tabular}
\caption{SMEFT Wilson coefficients truncated up to order $\mathcal{O}(\xi^3)$~\label{tab:SMEFT_couplings_xi}}
\end{center}
\end{table}

\section{Conclusion  and  Discussion~\label{sec:concl}}
In this work, we presented the HEFT matching to the RHTE . Inspired by the experimental constraint on the amount of custodial symmetry violation, we suggest using the ratio between the VEVs of the BSM scalar and doublet ($\xi\equiv v'/v$) as an expansion parameter in scalar sector extensions of the SM, like the RHTE, due to its tininess. With a certain input parameter set choice, we show that the masses of the BSM states obey the scaling $\sim\xi^{-1}$, which implies that they are decoupling. With the help of the non-linear representation proposed in Ref.~\cite{Song:2024kos}, we obtained the HEFT matching results up to $\mathcal{O}(\xi^3)$ for the first time, which can systematically be extended to higher orders. The matching equations for the parameters of the HEFT Lagrangian were discussed and given analytically.

We then investigated how accurately the HEFT matching reproduces the RHTE results in the tree-level scatterings $hh\rightarrow hh$, $WW\rightarrow hh$ and $ZZ\rightarrow hh$ compared to the SMEFT. Generally this is a non-trivial question due to the different power counting rules used in the two EFTs. We found that the HEFT at $\mathcal{O}(\xi^2)$ reproduces the RHTE results in all these scatterings with our choice of input parameters and power counting rules for two EFTs ($1/\Lambda$ for SMEFT and $\xi$ for HEFT). Especially at low collision energy, the convergence to the RHTE can be significantly improved if higher orders of $\xi$'s effects are included in the HEFT while the second order of the SMEFT expansion is not enough. We also show an explicit failure of the SMEFT in both a phase transition occurring and the characteristic scale changing in UV theory. The phase transition is closely related to the fundamental symmetry that should be used to build the EFT, which is already well known. That the characteristic scale in UV theory could be different in different regions of the parameter space is only explicitly pointed out here. The matching to the SMEFT in this scenario should be carefully reconsidered. Finally, we discussed that the naive method to obtain a ``consistent expansion'' in SMEFT to the one in HEFT could be problematic unless the decoupling limit can be explicitly made. This reveals again that the choice of the UV parameter set is important when one performs the EFT matchings, and the expansions are generally incommutable because of some singularity behaviors in the conversion functions. We believe that it is still possible to parameterize the BSM effects in the form of the SMEFT operators, though diagram approach should be applied. However, this causes a series problem that how we use the EFTs correctly in the aspect of the experimentalists. Since one can use either EFTs to parameterize the possible deviations observed in the near future experiments, the inverse map problem from the experimental data to the UV models should be carried out correctly when one uses an EFT description. As we show explicitly, the fundamental properties of the UV theory only show up in the correlations among the operators with different power counting orders in the EFT. Given the limited precisions of the future experiments, it is still very difficult to use just one EFT (SMEFT or HEFT) to uncover the underlying UV models. In practice, some methods to quantify the quality of the EFT Wilson coefficients in different EFTs should be proposed from the pure bottom-up view.

Several directions of future work are open. It would be particularly interesting to directly calculate the one-loop matching results to the HEFT in our framework, especially to figure out how the $SU(2)_L$ symmetry is restored in the operator level. UV models other than the RHTE, which have a non-vanishing scalar VEV contributing to the custodial violation, could also be explored. Also interesting would be the study of inverse map of a certain observed anomaly to the UV theories via two different EFT approaches, and whether some unique UV models can be obtained or not. 

\acknowledgments
The work of H.S. is supported by IBS under the project code, IBS-R018-D1. X.W. is supported by the National Science Foundation of China under Grants No. 11947416.

\appendix
\section{Field redefinition of the Higgs field and the HEFT matching results in a non-redundant basis}
The kinetic term for $h$ has acquired the form
\begin{align}
    \mathcal{L}_{h, kin}=\frac{1}{2}\mathcal{K}(h)D_\mu h D^\mu h
\end{align}
where
\begin{align}
    \mathcal{K}(h)=1+c_1\frac{h}{\vev}+c_2\frac{h^2}{\vev^2}+\cdots
\end{align}
The field redefinition
\begin{align}
    \Tilde{h}=\int_0^h\sqrt{\mathcal{K}(s)}ds=h\left(1+\frac{c_1}{4}\frac{h}{\vev}-\frac{c_1^2-4c_2}{24}\frac{h^2}{\vev^2}+\frac{c_1^3-4c_1 c_2+8c_3}{64}\frac{h^3}{\vev^3}+\cdots\right)
\end{align}
\begin{align}
    h=\Tilde{h}\left(1-\frac{c_1}{4}\frac{\Tilde{h}}{\vev}+\frac{c_1^2-c_2}{6}\frac{\Tilde{h}^2}{\vev^2}-\frac{7c_1^3-13c_1 c_2+6c_3}{48}\frac{\Tilde{h}^3}{\vev^3}+\cdots\right)
\end{align}
Eliminating $h$ in the EFT Lagrangian in favor of $\Tilde{h}$ and dropping the tilde in the end, the Wilson coefficients takes
\begin{table}[h!]
\centering
\begin{tabular}{c|cc}
\hline\hline
$\mathcal{O}(\xi)$ & $\xi^2$ & $\xi^3$ \\
\hline\hline
$\Delta\kappa_3$ & $-2\left(2-\frac{Z_3}{Z_1}\right)$ & $\frac{2Z_3}{Z_1^2}\frac{Y_3}{\vev}-\frac{4(12Z_1^2-7Z_1 Z_3+Z_3^2)}{Z_1}\frac{\vev}{Y_3}$ \\
$\Delta\kappa_4$ & $-4\left(\frac{22}{3}-\frac{3Z_3}{Z_1}\right)$ & $\frac{12Z_3}{Z_1^2}\frac{Y_3}{\vev}-\frac{8(4Z_1-Z_3)(41Z_1-15Z_3)}{3Z_1}\frac{\vev}{Y_3}$ \\ \hline
$\Delta a$ & $4$ & $4(4Z1-Z3)\frac{\vev}{Y_3}$ \\
$\Delta b$ & $16$ & $40(4Z1-Z3)\frac{\vev}{Y_3}$ \\ \hline
$\Delta\alpha$ & $2$ & $0$ \\
$\Delta a^\slashed{C}$ & $8$ & $8(4Z_1-Z_3)\frac{\vev}{Y_3}$ \\
$\Delta b^\slashed{C}$ & $12$ & $36(4Z_1-Z_3)\frac{\vev}{Y_3}$ \\ \hline\hline
$m_h^2$ & \multicolumn{2}{c}{$2\vev^2 Z_1 - 2\vev Y_3 \xi - 4v^2(4Z_1-Z_3)\xi^2 - 2\left[(4Z_1-Z_3)^2\vev^2-6Y_3^2\right]\frac{\vev}{Y_3}\xi^3$} \\ \hline
\end{tabular}
\caption{The Wilson coefficients of the HEFT $\mathcal{O}(p^2)$ operators and the squared mass of the Higgs $m_h^2$ for the triplet model in the standard non-redundant HEFT basis. All the couplings are shown up to $\mathcal{O}(\xi^3)$.~\label{tab:HEFT_couplings_norm}}
\end{table}

The leading terms of the Wilson coefficients of the $p^4$ operators do not change under the above field redefinition.

\section{Comments on 1-loop matching}
We present the complete UV Lagrangian in the following form
\begin{align}
    \mathcal{L}_{\rm RHTE}\supset&\frac{1}{2}\partial_\mu h\partial^\mu h-\frac{1}{2}m_h^2 h^2 - d_1 h^3 - z_1 h^4+\frac{1}{4}\left(2v_\Sigma^2-4v_\Sigma\sgam h+2\sgam^2 h^2\right)V_\mu^3 V^\mu_3 \nonumber \\
    &-\frac{1}{4}\left[v_H^2+4v_\Sigma^2+2\left(v_H\cgam-4v_\Sigma\sgam\right)h+\left(\cgam^2+4\sgam^2\right) h^2\right]\braket{V_\mu V^\mu} \label{eq:L_lfs} \\
    &+\left[-d_2 h^2-z_2 h^3-\frac{1}{2}\left(v_H\sgam+4v_\Sigma\cgam-3\sgam\cgam h\right)\braket{V_\mu V^\mu}+\left(v_\Sigma\cgam-\sgam\cgam h\right)V_\mu^3 V^\mu_3\right]K \nonumber \\
    &+\left[\left(v_\Sigma-\sgam h\right)V_\mu^1 V^\mu_3-i\left(\sgam+v_\Sigma\cgam/v_H\right)V_\mu^2\left(h D^\mu-D^\mu h\right)\right]\phi_1 \nonumber \\
    &+\left[\left(v_\Sigma-\sgam h\right)V_\mu^2 V^\mu_3+i\left(\sgam+v_\Sigma\cgam/v_H\right)V_\mu^1\left(h D^\mu-D^\mu h\right)\right]\phi_2 \label{eq:L_hfl} \\
    &+\frac{1}{2}\begin{pmatrix}
        K & \phi_1 & \phi_2
    \end{pmatrix}\mathcal{X}\begin{pmatrix}
        K \\ \phi_1 \\ \phi_2
    \end{pmatrix} \label{eq:L_hfq} \\
    & - d_4 K^3 - d_6 K\phi^+\phi^- - z_4 h K^3 - z_5 K^4 - z_7 hK\phi^+\phi^- - z_8 K^2\phi^+\phi^- - z_9\left(\phi^+\phi^-\right)^2 \label{eq:L_hfHO}
\end{align}
where we define
\begin{gather}
    \mathcal{X}=\begin{pmatrix}
        \substack{-\partial^2-m_K^2-2d_3 h-2z_3 h^2 \\ -\left(1+3\cgam^2\right)\braket{V_\mu V^\mu}/2+\cgam^2 V_\mu^3 V^\mu_3} & \substack{\cgam V_\mu^1 V^\mu_3+2i\left(\cgam-v_\Sigma\sgam/v_H\right)V_\mu^2 D^\mu} & \substack{\cgam V_\mu^2 V^\mu_3-2i\left(\cgam-v_\Sigma\sgam/v_H\right)V_\mu^1 D^\mu} \\
        \substack{\cgam V_\mu^3 V^\mu_1-2i\left(\cgam-v_\Sigma\sgam/v_H\right)V_\mu^2 D^\mu} & \substack{-\left(1+4v_\Sigma^2/v_H^2\right)D^2-m_{\phi^\pm}^2-d_5 h-z_6 h^2 \\ -2\left(1+v_\Sigma^2/v_H^2\right)\braket{V_\mu V^\mu}+V_\mu^1 V^\mu_1} & \substack{V_\mu^1 V^\mu_2+2i(1+2v_\Sigma^2/v_H^2)V_\mu^3 D^\mu} \\
        \substack{\cgam V_\mu^3 V^\mu_2+2i\left(\cgam-v_\Sigma\sgam/v_H\right)V_\mu^1D^\mu} & \substack{V_\mu^2 V^\mu_1-2i(1+2v_\Sigma^2/v_H^2)V_\mu^3 D^\mu} & \substack{-\left(1+4v_\Sigma^2/v_H^2\right)D^2-m_{\phi^\pm}^2-d_5 h-z_6 h^2 \\ -2\left(1+v_\Sigma^2/v_H^2\right)\braket{V_\mu V^\mu}+V_\mu^2 V^\mu_2}
    \end{pmatrix} \\
    D^\mu_{ij}\equiv\partial^\mu\delta_{ij}-g' B^\mu\epsilon_{ij},\text{ with } i, j\in{1, 2} \\
    V_\mu^i=\braket{U^\dagger D_\mu U\sigma_i}
\end{gather}
Note that following the standard CDE matching procedure, we classify the Lagrangian terms into pure light fields terms Eq.~\ref{eq:L_lfs}, terms linear in heavy fields Eq.~\ref{eq:L_hfl}, quadratic terms in heavy fields Eq.~\ref{eq:L_hfq}, and higher-order terms in heavy fields Eq.~\ref{eq:L_hfHO}. In principle, one can use the CDE method to integrate out the heavy states $K, \phi_{1, 2}$ at the 1-loop level to obtain the EFT results.

Several aspects are worth mentioning here. The first one is that, for compactness, we write the Lagrangian in terms of $V_\mu^i$ instead of Goldstone matrix $U$. However, it should be noted that in the path integral formulation, it is the field $U$ being integrated over its field configurations. If one treats $V_\mu^i$ as the fundamental fields, which is a very straightforward treatment since $V_\mu^i$ can be considered as the longitudinal modes of the massive gauge bosons, it seems that there is no kinetic term of $V_\mu^i$ in the above expression. So this field redefinition should generate a non-trivial Jacobian factor. Second, the covariant derivative acting on the heavy fields (therefore also on the field $V_\mu^i$) at most depends on the hypercharge gauge field. Naively, one may guess that only terms involving $B_{\mu\nu}$ can be generated but no terms containing $W_{\mu\nu}$. Or in another way to say, the form of the Lagrangian given above is not invariant under the $SU(2)_L$ symmetry, so it is not clear how the $SU(2)_L$ symmetry is restored in the HEFT after matching. Therefore, We leave the complete 1-loop to the future work.

\bibliographystyle{JHEP}
\bibliography{reference}

\end{document}